\documentclass[11pt,a4paper]{article}
\pdfoutput=1
\usepackage{jheppub}

\usepackage{multirow, graphicx,amssymb,url,mathrsfs,amsmath}
\usepackage{wrapfig,boxedminipage,setspace,subfigure,epsfig}
\usepackage{amsxtra,amstext,latexsym,dsfont,amsfonts}
\usepackage{color,eucal}
\usepackage[dvipsnames]{xcolor}
\usepackage{float}

\newcommand{\dd}{\mathrm{d}}

\newcommand{\grad}{\nabla}

\newcommand{\yj}{\xi}

\newcommand{\dja}{\varphi_0}

\title{Thermal diffusivity and butterfly velocity in anisotropic Q-Lattice models}

\author[a]{Dujin Ahn,}
\author[a]{Yongjun Ahn,}
\author[a]{Hyun-Sik Jeong,}
\author[a]{Keun-Young Kim,}
\author[b]{Wei-Jia Li}
\author[a]{and Chao Niu}

\emailAdd{dujinahn@gmail.com}
\emailAdd{yongjunahn619@gmail.com}
\emailAdd{hyunsik@gist.ac.kr}
\emailAdd{fortoe@gist.ac.kr}
\emailAdd{weijiali@dlut.edu.cn}
\emailAdd{chaoniu09@gmail.com}

\affiliation[a]{School of Physics and Chemistry, Gwangju Institute of Science and Technology,
	Gwangju 61005, Korea}
\affiliation[b]{Institute of Theoretical Physics, School of Physics, Dalian University of Technology, Dalian 116024,
China}

\abstract{
By using a holographic method we study a relation between the thermal diffusivity ($D_T$) and two quantum chaotic properties, Lyapunov time ($\tau_L$) and butterfly velocity ($v_B$) in strongly correlated systems.  It has been shown that $D_T/(v_B^2 \tau_L)$ is universal in some holographic models as well as condensed matter systems including the Sachdev-Ye-Kitaev (SYK) models. We investigate to what extent this relation is universal in the Q-lattice models with infrared (IR) scaling geometry, focusing on the effect of spatial anisotropy. Indeed it was shown that $\mathcal{E}_i := D_{T,i}/(v_{B,i}^2 \tau_L)$ ($i=x,y$) is determined only by some scaling exponents of the IR metric  in the low temperature limit regardless of the matter fields and ultraviolet   data. Inspired by this observation, in this work, we find the concrete expressions for $\mathcal{E}_i$ in terms of the critical dynamical exponents $z_i$ in each direction. By analyzing the IR scaling geometry we identify the allowed scaling parameter regimes, which enable us to compute the allowed range of $\mathcal{E}_i$.
We find the lower bound of $\mathcal{E}_i$ is always $1/2$, which is not affected by anisotropy, contrary to the $\eta/s$ case. However, there may be an upper bound determined by anisotropy.
 }

\begin{document}

\maketitle

\section{Introduction} \label{sec:1}

Strongly correlated electron systems are characterized by exotic (`strange') properties in contrast to  weakly interacting systems.  Interestingly enough, some exotic properties show a remarkable degree of universality~\cite{Hartnoll:2016apf}.
For example, resistivity ($\rho$) is observed to be linear in temperature ($T$), $ \rho \sim T$, universally in various strange metals such as cuprates, pnictides and heavy fermions.
It is in contrast to $ \rho \sim T^2$ which can be explained by the Fermi liquid theory for metals with weakly  interacting electrons. The strange metal state may undergo a phase transition to the high temperature superconducting state, where another universal property, Home's law~\cite{Homes:2004wv, Zaanen:2004aa,Erdmenger:2015qqa}, has been observed.
It is a relation between three quantities: the superfluid density at zero temperature $\rho_s(T=0)$,  the critical temperature ($T_c$), and the DC electric conductivity right above the critical temperature ($\sigma_{\mathrm{DC}}(T_c)$). The Homes' law states that $\rho_s(T=0)/( \sigma_{\mathrm{DC}}(T_c) T_c)$ is universal, which means it is independent of the components and structures of superconducting materials.

While such interesting properties in strongly correlated systems are difficult to analyze theoretically,
the holographic methods (or the gauge/gravity duality)~\cite{Zaanen:2015oix, Ammon:2015wua,Hartnoll:2016apf}  have provided novel and effective tools to investigate them. It maps strongly correlated systems to corresponding classical gravitational systems in higher dimensional spacetime, so `holographic'. For example, for linear-$T$-resistivity, a lot of achievements including methodologies are reviewed in \cite{Hartnoll:2016apf}. For the Homes' law, see \cite{Erdmenger:2015qqa,Kim:2015dna,Kim:2016hzi,Kim:2016jjk}. In general, universality in strongly correlated systems is related to the universal nature of the black hole horizon. Therefore, investigating as many universal properties in strongly correlated systems as possible will be helpful in understanding the black hole physics better. This will again back-react to our understanding of strongly correlated systems. 

In this paper, we investigate another universal property regarding the thermal diffusivity and two quantum chaotic properties, butterfly velocity and Lyapunov time,  from the holographic perspective.
The thermal diffusivity ($D_T$) is defined by
\begin{equation} \label{kuy}
D_T := \frac{\kappa}{c_\rho} \,,
\end{equation}
where $c_\rho$ is the specific heat at finite density and $\kappa$ is the thermal conductivity in the open circuit condition, i.e. at zero electric current. It was proposed that the thermal diffusivity has an interesting connection to the quantum chaos property as follows\footnote{This kind of relation was
first motivated by the charge diffusivity and its relation to the linear-$T$-resistivity~\cite{Hartnoll:2014lpa, Blake:2016wvh, Blake:2016sud}. However, it turned out the relation \eqref{chaos111} for charge diffusivity does not hold in many models, for example, striped holographic matter \cite{Lucas:2016yfl}, the SYK model~\cite{Davison:2016ngz}, higher derivative models~ \cite{Baggioli:2016pia} and the Gubser-Rocha model~\cite{Kim:2017dgz}. }:
\begin{equation} \label{chaos111}
D_T = \mathcal{E} v_B^2 \tau_L  \,,
\end{equation}
where $v_B$ is the butterfly velocity which describes the speed at which chaos propagates in space~\cite{Sekino:2008he, Shenker:2013pqa, Roberts:2014isa, Maldacena:2015waa, Blake:2016wvh, Roberts:2016wdl, Blake:2016sud, Ling:2016ibq, Alishahiha:2016cjk,Qaemmaqami:2017jxz,Jahnke:2017iwi}, and
$\tau_L$ is the Lyapunov time which measures the rate at which chaos grows in time.
$\mathcal{E}$ is constant of order one.
It was shown~\cite{Maldacena:2015waa} that there is a universal lower bound for
$\tau_L$: 
\begin{equation} \label{lklk}
\tau_L \ge \frac{1}{2\pi }\frac{\hbar}{k_B T} =:\frac{1}{2\pi} \tau_P \,,
\end{equation}
where the timescale ($\tau_P$) was introduced in \cite{Sachdev:2011cs, Zaanen:2004aa} as the `Planckian' dissipation time scale, which is the shortest possible time scale for dissipation.
This time scale was observed in the scattering rates of materials having a linear $T$ resistivity~\cite{Bruin804} and in the thermal diffusivity~\cite{Zhang:2016aa}.
The Lyapunov time saturates the bound in holographic theories with Einstein gravity. While the connection \eqref{chaos111} between transport properties and chaos was first proposed in the holographic models, it has been also observed in condensed matter systems \cite{Aleiner:2016aa, Swingle:2016aa, Patel:2016aa, Zhang:2016aa}.

The relation \eqref{chaos111} is shown to be {universal} in several cases.
It was shown that, in a class of holographic model with a scaling infra-red (IR) geometry characterized by critical exponents such as dynamical critical exponent ($z$),  hyperscaling violating exponent ($\theta$) or charge anomalous parameter ($\zeta$)~\cite{Gouteraux:2014hca}, $\mathcal{E}$ is  a function only of a dynamical critical exponent $z \ne 1$
\begin{equation} \label{uytr0}
\mathcal{E} = \frac{1}{2}\frac{z}{z-1} \,,
\end{equation}
at zero density in the low temperature limit, independently of other critical exponents, momentum relaxation strength and UV data~\cite{Blake:2016sud}.
Recently this analysis was extended to finite density or magnetic field case and $\mathcal{E}$ is shown to be  independent of charge density and magnetic field too~\cite{Blake:2017qgd}.
More evidences for \eqref{chaos111} have been reported in holographic models that flow to AdS$_2 \times R^d$ fixed points in the IR~\cite{Blake:2016jnn} and in the higher derivative model~\cite{Baggioli:2016pia, Wu:2017mdl,Baggioli:2017ojd}.  At finite density, there is an issue in defining the thermal diffusivity because of its mixing with the charge diffusivity. However, it was shown in~\cite{Kim:2017dgz} that the mixing effect becomes negligible in the incoherent regime (i.e. the regime of strong momentum relaxation) and $\mathcal{E}$ becomes universal in that regime even at finite density. More interestingly,
while the relation \eqref{chaos111} was first proposed in the holographic models, it has been also observed in condensed matter systems \cite{Aleiner:2016aa, Swingle:2016aa, Patel:2016aa, Zhang:2016aa, Bohrdt:2016vhv,Werman:2017abn} including the Sachdev-Ye-Kitaev (SYK) models~\cite{Gu:2016oyy, Davison:2016ngz, Jian:2017unn}.

In this paper, we want to study the effect of spatial anisotropy\footnote{For the effect of spatial anisotropy on shear viscosity, see \cite{Giataganas:2017koz}} on the universality of $\mathcal{E}$, more precisely, two $\mathcal{E}$'s, one for the $x$-direction ($\mathcal{E}_x$) and one for the $y$-direction ($\mathcal{E}_y$).
This question was already addressed in \cite{Blake:2017qgd}, where it was shown that both $\mathcal{E}_x$ and $\mathcal{E}_y$ are determined only by some scaling parameters ($u_1, v_1, v_2$) near horizon:
\begin{equation} \label{uytr1}
	\mathcal{E}_x  = \frac{u_{1} - 1}{u_{1} - 2v_{1}}, \qquad \mathcal{E}_y =  \frac{u_{1} - 1}{u_{1} - 2v_{2}} \,.
\end{equation}
Here, we continue the analysis and find the expression in terms of the dynamical exponents $z_i$ for $i$-direction ($i=x,y$):
\begin{equation} \label{uytr2}
	\mathcal{E}_x  = \frac{1}{2}\frac{z_x}{z_x-1}  , \qquad \mathcal{E}_y  = \frac{1}{2}\frac{z_y}{z_y-1}  \,,
\end{equation}
which clearly show  universality in terms of physical parameters and the effect of anisotropy. $\mathcal{E}_{x}$ and $\mathcal{E}_{y}$ do not depend on other critical exponents ($\theta$, $\zeta$) and charge density $\rho$.  This universality is due to  nontrivial cancellations between
three quantities $\{\kappa, c_\rho, v_{B}\}$, all of which depend on many other parameters including UV data. 

So far we discussed the universality of $\mathcal{E}_i$ in the sense that $\mathcal{E}_i$  is independent of many IR parameters as well as UV data. However, it is also interesting to see if there is any universal lower or upper\footnote{The existence of the upper bound was proposed in \cite{Gu:2017ohj, Hartman:2017hhp}} {\it bound} of $\mathcal{E}_i$. For example, in the case of \eqref{uytr0} and \eqref{uytr2} it amounts to asking if there is any universal bound of dynamical critical exponents $z_i$.
For the isotropic IR scaling geometry it was shown that $z>1$~\cite{Gouteraux:2014hca}, which implies\footnote{Mathematically $z<0$ is allowed, but we do not consider it here: it is not physical because $\omega \sim k^z$~\cite{Ling:2016yxy}. }
\begin{equation} \label{qrwe}
\frac{1}{2} \le \mathcal{E} \,,
\end{equation}
where $\mathcal{E}$ saturates its minimum value $\mathcal{E} = 1/2$ as $z \rightarrow \infty$. 
For the theories that flow to AdS$_2 \times R^d$ IR fixed points, $\mathcal{E}$ depends only on the leading irrelevant mode and $1/2 < \mathcal{E} \le 1$ where $\mathcal{E} = 1$ if the leading deformation is  a dilatonic mode~\cite{Blake:2016jnn}. It matches the value in extended  SYK models~\cite{Gu:2016oyy,Davison:2016ngz}. However, it was reported that $\mathcal{E}$ may not have a universal lower bound in an {\it inhomogeneous} SYK model~\cite{Gu:2017ohj} or in a higher derivative gravity theory~\cite{Li:2017ncu}. 

Here we investigate if the range \eqref{qrwe} for the isotropic scaling geometry can be affected by {\it anisotropy}.  
It was motivated by a series of works regarding the universal bound of sheer viscosity to entropy density ratio ($\eta/s$) so called the KSS(Kovtun, Starinets and Son) bound, $1/4\pi$~\cite{Kovtun:2004de}. It can be lowered further by the higher derivative gravity~\cite{Brigante:2007nu,Brigante:2008gz}. However, it can even vanish by anisotropy at zero temperature\footnote{Momentum relaxation gives similar results at zero temperature~\cite{Hartnoll:2016tri,Alberte:2016xja,Burikham:2016roo,Ling:2016ien}.}\cite{Rebhan:2011vd,Mamo:2012sy,Jain:2015txa,Jain:2014vka}.
In our anisotropic model, we find that the lower bound of $\mathcal{E}_i$ is always $1/2$ regardless of anisotropy but the upper bound  of $\mathcal{E}_i$ may depend on anisotropy. In other words, $z_i$ depends on anisotropy.

This paper is organized as follows.
In section \ref{sec:2}, we introduce the `Q-lattice model' or the Einstein-Maxwell-Dilaton theory coupled to `axion' fields. By assuming specific couplings in the IR region, we obtain the IR scaling solutions described by four scaling parameters, and classify the solutions according to the IR relevance of gauge field and axion field. We also analyze the allowed parameter region of the solutions by requiring some physical conditions.
In section \ref{sec:4} we study the thermal diffusivity, butterfly velocity, and their universality based on the results obtained in section \ref{sec:2}. 
In section \ref{sec:5}, we conclude.

\section{IR analysis for anisotropic Q-Lattice models}  \label{sec:2}
Let us consider the `Q-lattice action'
\begin{equation}
\begin{split} \label{EMDA}
S &=  \int\dd^{p+1}x   \sqrt{-g}   \left ( R  + \mathcal{L}_{m}  \right) \,, \\
 & \quad \mathcal{L}_{m}  \equiv -  \frac{1}{2}(\partial\varphi)^2+ V(\varphi)  -  \frac{1}{4} Z(\varphi) F^2 -\frac{1}{2}W_1(\varphi)\sum_{i=1}^{p-2}(\partial \chi_{i})^2  -\frac{1}{2}W_2(\varphi)(\partial \chi_{p-1})^2 \,,
\end{split}
\end{equation}
which is the Einstein-Maxwell-Dilaton theory coupled to `Axion' fields $\chi_i$ ($i=1, \dots, p-1$). This model is also called the EMD-Axion action.  We introduce the axions as many as spatial dimensions and every axion may have different coupling in general, say $W_i(\varphi)$. However, for simplicity we introduce anisotropy minimally by two couplings  $W_1$ and $W_2$.   We will further assume the axion fields have the form
\begin{equation}
\chi_{i}=k_{1} x_{i} \ \ (i=1, \dots, p-2)\,, \qquad \chi_{p-1}=k_{2} y \,,
\end{equation}
to break translational symmetry. Here we introduced another anisotropy by $k_1$ and $k_2$.   In summary,
we introduced two kinds of anisotropy: i) in the action, $W_1$ and $W_2$  ii) in the solution $k_1$ and $k_2$.

The action yields the following Einstein equations:
%
%
\begin{align} \label{eommaster}
\begin{split}
 R_{\mu\nu} &= T_{\mu\nu} - \frac{1}{p-1}g_{\mu\nu}T\\
&=\frac{1}{2}\partial_{\mu}\varphi\partial_{\nu}\varphi
+\frac{W_{1}(\varphi)}{2}\sum_{i=1}^{p-2}\partial_{\mu}\chi_{i}\partial_{\nu}\chi_{i}+\frac{W_{2}(\varphi)}{2}\partial_{\mu}\chi_{p-1}\partial_{\nu}\chi_{p-1}+\frac{Z(\varphi)}{2}F_{\mu}{^\rho}F_{\nu\rho}\\
&\quad -\frac{Z(\varphi)F^2}{4(p-1)}g_{\mu\nu}-\frac{V(\varphi)}{p-1}g_{\mu\nu} \,, \\
\end{split}
\end{align}
where $T_{\mu\nu} = -\frac{1}{\sqrt{-g}}\frac{\delta(\sqrt{-g}\mathcal{L}_{m})}{\delta g^{\mu\nu}}$ and $T = g^{\mu\nu}T_{\mu\nu}$ \,. The Maxwell equation, scalar equation, and axion equation read
\begin{align} \label{eommaster2}
\begin{split}
&\grad_{\mu}(Z(\varphi)F^{\mu\nu}) = 0 \,, \\& \square\varphi+V'(\varphi)-\frac{1}{4}Z'(\varphi)F^2-\frac{1}{2}W_{1}'(\varphi)\sum_{i=1}^{p-2}(\partial\chi_{i})^2-\frac{1}{2}W_{2}'(\varphi)(\partial\chi_{p-1})^2 =0
  \,, \\
&\grad_{\mu}(W_{1}(\varphi)\grad^{\mu}\chi_{i}) =0 \,,  \qquad  \grad_{\mu}(W_{2}(\varphi)\grad^{\mu}\chi_{p-1}) =0 \,.
\end{split}
\end{align}
By considering the following homogeneous (all functions are only functions of $r$) ansatz
\begin{align} \label{backmaster}
\begin{split}
&\dd s^2=-D(r)\dd t^2+B(r)\dd r^2+C_{1}(r)\sum_{i=1}^{p-2}\dd x_{i}^{2}+C_{2}(r)\dd y^{2} \,,\\
&\varphi=\varphi(r) \,, \quad A=A_t(r) \dd t \,, \quad \chi_{i}=k_{1} x_{i} \,, \quad \chi_{p-1}=k_{2} y \,,
\end{split}
\end{align}
we obtain the Einstein equations 
 %
\begin{align}
&  \!  0=\frac{Z(p-2)A_{t}'^{2}}{(p-1)D}+\frac{2BV}{p-1}+\frac{B'D'}{2BD} -\frac{(p-2)D'C'_{1}}{2DC_{1}}-\frac{D'C'_{2}}{2DC_{2}}+\frac{D'^{2}}{2D^{2}}-\frac{D''}{D}\,,  \label{eom11} \\
&  \!  0 =\varphi'^{2}-\frac{(p-2)C'^{2}_{1}}{2C^{2}_{1}}-\frac{C'^{2}_{2}}{2C^{2}_{2}}-\frac{D'}{2D}\left(\frac{(p-2)C'_{1}}{C_{1}}+\frac{C'_{2}}{C_{2}}\right)-\frac{B'}{2B}\left(\frac{(p-2)C'_{1}}{C_{1}}+\frac{C'_{2}}{C_{2}}\right)\nonumber \\
  & \qquad +\frac{(p-2)C''_{1}}{C_{1}}+\frac{C''_{2}}{C_{2}}\,, \label{eom12} \\
&  \! 0=\frac{W_{1} k^{2}_{1}B}{C_{1}}-\frac{2BV}{p-1}+\frac{ZA_{t}'^{2}}{(p-1)D}+\frac{C'_{1}}{2C_{1}}\left(\frac{D'}{D}-\frac{B'}{B}\right)+\frac{(p-4)C'^{2}_{1}}{2C^{2}_{1}}+\frac{C'_{1}C'_{2}}{2C_{1}C_{2}}+\frac{C''_{1}}{C_{1}}\,, \label{eom13} \\
& \! 0=\frac{W_{2} k^{2}_{2}B}{C_{2}}-\frac{2BV}{p-1}+\frac{ZA_{t}'^{2}}{(p-1)D}+\frac{C'_{2}}{2C_{2}}\left(\frac{D'}{D}-\frac{B'}{B}\right)-\frac{C'^{2}_{2}}{2C^{2}_{2}}+\frac{(p-2)C'_{1}C'_{2}}{2C_{1}C_{2}}+\frac{C''_{2}}{C_{2}}\,, \label{eom14}
\end{align}
which come from the equations corresponding to $R_{tt}, R_{rr}, R_{xx}$, and $R_{yy}$ in \eqref{eommaster} respectively.
The prime $'$ denotes the derivative with respect to $r$. The Maxwell equation and scalar equation are reduced to
\begin{align}
&0=\left[Z\frac{C_{1}^{\frac{p-2}{2}}C_{2}^{\frac{1}{2}}}{\sqrt{BD}}A_{t}'\right]'\,, \label{max123} \\
&0=-\frac{W_{1, \varphi}k^{2}_{1}(p-2)B}{2C_{1}}-\frac{W_{2, \varphi}k^{2}_{2}B}{2C_{2}}+\frac{Z_{,\varphi}A_{t}'^{2}}{2D}+BV_{,\varphi}-\frac{B'\varphi'}{2B} \nonumber \\ \label{axax}
  &\quad \ \ +\left(\frac{(p-2)C'_{1}}{2C_{1}}+\frac{C'_{2}}{2C_{2}}\right)\varphi' +\frac{D'\varphi'}{2D}+\varphi'' \,,
\end{align}
and the axion equations are satisfied trivially.

\subsection{General structure of the IR solutions}

In this paper, we are mainly interested in the scaling geometry at IR, where $\varphi$ runs logarithmically and the dilaton couplings are approximated as
\begin{align} \label{IRcoupling}
Z(\varphi) \sim e^{\gamma \varphi} \,, \qquad V(\varphi) \sim V_{0}e^{-\delta \varphi}\,, \qquad W_{1}(\varphi) \sim e^{\lambda_{1}\varphi}\,,\qquad W_{2}(\varphi) \sim e^{\lambda_{2}\varphi} \,.
\end{align}
Here we introduce parameters $(\gamma, \delta, \lambda_1, \lambda_2, V_0)$, which we call `action-parameters'.  For $Z, W_1, W_2$ we do not introduce the coefficients because they can be absorbed into the gauge field and axions. 
To analyze the IR solution, we will plug the IR couplings  \eqref{IRcoupling} into \eqref{eom11}-\eqref{axax} and 
assume that the IR solutions are written as
\begin{equation} \label{generalsol}
\begin{split}
& \dd s^2=r^{\frac{2\theta}{p-1}}\left[-\frac{\dd t^2}{r^{2z}}+\frac{L_r^{2}\dd r^2
}{r^2}+\frac{L_{1}^{2}\,\sum\limits_{i=1}^{p-2} \dd x_{i}^2}{r^{2\xi_x}}+\frac{L_{2}^{2}\,\dd y^2}{r^{2{\xi_y}}}\right]\,,   \\
&  {\varphi}={\dja}\log r  \,,  \quad A=a_{0}\,r^{\zeta-z}\,\dd t \,,  \quad \chi_{i}=k_{1} x_{i} \,, \quad \chi_{p-1} =k_{2} y \,,
  \end{split}
\end{equation}
in terms of  `exponents' ($z, \theta, \xi_x, \xi_y, \zeta$) and `coefficients'  ($\dja, a_0, L_r, L_1, L_2, k_1, k_2$).  We will call all of them `solution-parameters', to explain their relations to the `action-parameters' which are the parameters in the action. For example, by the equations of motion, the exponent-solution-parameters $(z,\theta, \xi_x, \xi_y,  \zeta)$  will be related to the action parameters $(\gamma, \delta$, $\lambda_1, \lambda_2)$.
Notice that this kind of scaling solutions \eqref{generalsol} are possible since the scalar is of the form $(e^\varphi = r^{\dja})$. 

  Some of solution-parameters are redundant and can be set to unity by coordinate transformations.   Depending on our purpose and perspective we may choose independent parameters without loss of generality. 
In this paper, we will choose the representation with 
\begin{equation}
\xi_x=1 \,, \quad  \xi_y = \yj \,,
\end{equation}
for an easy comparison with the isotropic results obtained in \cite{Gouteraux:2014hca}.  We will mostly fix $L_1 =L_2 =1$  but sometimes we find it more convenient to keep $L_1$ and $L_2$ unfixed. 

The metric is parameterized in a way to identify four critical exponents. There are two dynamical critical exponents 
\begin{equation}
 z_x :=  \frac{z}{\xi_x} = z \,, \qquad z_y :=  \frac{z}{\xi_y} = \frac{z}{\yj} \,.
\end{equation}
They desrcibe the anisotropy between time and space $x_i$, and time and space $y$ respectively, where $\yj$ quantifies the anisotropy between $x_i$-space and $y$-space.  A hyperscaling violating exponent $\theta$ measures how much the scale invariance of the metric is violated and has something to do with the anomalous dimension of the field theory energy density.  $\zeta$ describing the anomalous scaling of the bulk Maxwell field is related to the anomalous dimension of the field theory charge density.

Furthermore, it turns out that the emblackening factor $f(r)$ 
\begin{equation} \label{emblack1}
f(r) =  1- \left(\frac{r}{r_h}\right)^{z+p-2-\theta+ \yj} \,,
\end{equation}
can be turned on ($\dd t^2 \rightarrow f \dd t^2$ and $\dd r^2 \rightarrow \dd r^2/f$ in \eqref{generalsol}) in all cases we consider in this paper. 

Some of the solutions parameters are fixed by the action parameters by the equations of motion but some of them are not fixed and remain free. However, the range of all parameters should be restricted by the following conditions. 
First, for the IR geometry to be well-defined, we require
\begin{equation} \label{con111}
\theta > (p-1)z \,, \quad \theta > p-1 \,,  \quad \theta > (p-1)\yj\,, \quad \theta >z+ p-2 +\yj \,,
\end{equation}
if the IR is located at $r \rightarrow 0$  or
\begin{equation} \label{con222}
\theta < (p-1)z \,, \quad \theta < p-1 \,,  \quad \theta < (p-1)\yj\,, \quad \theta <z+ p-2 +\yj \,,
\end{equation}
if the IR is located at $r \rightarrow \infty$.  The first three inequalities of \eqref{con111} and \eqref{con222} come from the condition that all metric components should vanish at the IR at zero $T$. The last inequalities come from the condition that the emblackening factor \eqref{emblack1} should vanish at the UV.  We also require\footnote{If we choose the representation $L_1=L_2=1$, we need to consider other reality conditions equivalent to them.}
\begin{equation}
L_r^2 > 0\,, \quad L_1^2 > 0 \,, \quad L_2^2 > 0 \,,
\end{equation}
and the specific heat  should be positive:
\begin{equation} \label{con7}
\frac{-2 \theta +(p-1)(\yj+1)}{(p-1) z} >0 \,,
\end{equation}
which can be read from the scaling of entropy, $S \sim T^{\frac{-2\theta +(p-1)(\yj+1)}{(p-1)z}}$.
If all of the above conditions are satisfied we have confirmed that the following null energy condition (NEC) is also satisfied:
\begin{align}
((p-1)-\theta ) ((p-1) (z-1)-\theta )-(\yj-1)  (2 (\yj-1) +(2-z) (p-1))\geq 0 \,, \\
(z-1) (-\theta +\yj +p+z-2)\geq 0 \,, \\
(z-1) (-\yj +z)\geq 0 \,.
\end{align}

We categorize the solutions according to the `relevance' of the axion and/or current, following \cite{Gouteraux:2014hca} for the easy comparison with the isotropic case therein.  
By `marginally relevant axion' we mean the axion parameter $k_1, k_2$ appear explicitly in the leading solutions and by `marginally relevant current' we mean $a_0$ appears explicitly in the leading solutions. By `irrelevant axion (current)'  
we mean $k_1, k_2$ $(a_0)$ do not appear explicitly in the leading solutions but they can appear in the sub-leading solutions. Therefore, we will consider eight classes as follows. 
%
%

%
\begin{itemize}
\item{class I: marginally relevant axions $\&$ current \qquad \qquad \qquad \ \ \ ($k_1 \ne 0,  k_2 \ne 0, a_0 \ne 0$) }
\item{class II: marginally relevant axions $\&$ irrelevant current \qquad \, ($k_1 \ne 0,  k_2 \ne 0, a_0 = 0$) }
\item{class III: irrelevant axions $\&$ marginally relevant current \qquad ($k_1 = 0,  k_2 = 0, a_0 \ne 0$) }
\item{class IV: irrelevant axions $\&$ current \qquad \qquad \qquad \qquad \qquad  \ \ ($k_1 = 0,  k_2 = 0, a_0 = 0$) }
\item{class I-i: mixed axions $\&$ marginally relevant current \qquad \qquad ($k_1 \ne 0,  k_2 = 0, a_0 \ne 0$) }
\item{class I-ii: mixed axions $\&$ marginally relevant current \qquad \ \ \, ($k_1 = 0,  k_2 \ne 0, a_0 \ne 0$) }
\item{class II-i: mixed axions $\&$  irrelevant current \qquad \qquad \qquad \ \ \, ($k_1 \ne 0,  k_2 = 0, a_0 = 0$) }
\item{class II-ii: mixed axions $\&$ irrelevant current \qquad \qquad \qquad \ \ \, ($k_1 = 0,  k_2 \ne 0, a_0 = 0$) }
\end{itemize}
Here, `mixed axions' means the axion in one direction is marginally relevant and the axion in the other direction is irrelevant. They reduce to four classes in \cite{Gouteraux:2014hca} in the isotropic limit.

Notice that the classification is based on the property of the leading solutions. We also should consider the deformation by the sub-leading solutions:
\begin{equation}
\Phi_i \rightarrow \Phi_i+ \epsilon_i r^{\beta_i} + \cdots  \,,
\end{equation}
where $\Phi_i$ denotes every leading order solution collectively and $\epsilon_i$ is a small parameter.
Therefore, $a_0=0$ does not mean zero density and $k_i=0$ does not mean no momentum relaxation in the $i$-direction because these parameters can appear in the sub-leading solutions.
If the axion is relevant, we may expect the momentum relaxation affects IR physics more strongly than the irrelevant axion cases.

\subsection{Marginally relevant axion}  \label{sec:21}

\subsubsection{Class I: marginally relevant current}

We assume that the classical solutions are written as
\begin{equation} \label{class1sol}
\begin{split}
& \dd s^2=r^{\frac{2\theta}{p-1}}\left[-\frac{\dd t^2}{r^{2z}}+\frac{L_r^{2}\dd r^2
}{r^2}+\frac{\sum\limits_{i=1}^{p-2} \dd x_{i}^2}{r^2}+\frac{\dd y^2}{r^{2\yj}}\right]\,,   \\
& {\varphi}={\dja}\log r \  \,,  \quad  A=a_{0}\,r^{\zeta-z}\,\dd t \,,  \quad \chi_{i}=k_{1} x_{i} \,, \quad \chi_{p-1} =k_{2} y \,,
  \end{split}
\end{equation}
where $\dja, a_0, k_1$, and $k_2$ are nonzero and $\zeta \ne z$. 

By the equations of motion, the `exponent' solution-parameters ($z, \yj, \theta,  \zeta$) may be expressed in terms of action-parameters ($\delta,  \lambda_1, \lambda_2, \gamma $) as
\begin{equation} \label{class11}
\begin{split}
 z&=\frac{2-(p-1)\delta^{2}+(p-2)\lambda^{2}_{1}+\lambda^{2}_{2}}{\lambda_{1}((p-1)\delta+(p-2)\lambda_{1}+\lambda_{2})}\,,  \\
 \theta &= - \frac{(p-1)\delta}{\lambda_{1}} \,, \quad \yj =\frac{\lambda_{2}}{\lambda_{1}}\,,  \quad \zeta =  \frac{\gamma-\delta}{\lambda_1} \,.
 \end{split}
\end{equation}
They are not all independent and there is a constraint between solution-parameters ($\theta, \zeta, \yj$)
\begin{equation} \label{con2}
\zeta  = -(p-2 -\theta) -\yj \equiv - (d_\theta-1) -\yj \equiv \zeta_I \,,
\end{equation}
which amounts to a relation between action-parameters:
\begin{equation} \label{con1}
 \gamma =  (2-p)\delta+(2-p)\lambda_{1}-\lambda_2 \,.
\end{equation}
The four `coefficient' parameters are solved as
\begin{equation}
\dja = -\frac{2}{\lambda_1} \,,
\end{equation}
and
\begin{align}
\begin{split}
&a_{0}^{2} = \\
& \frac{2 \left(k_1^2 (p-1) \left(\delta  \lambda _2+\lambda _2^2+\lambda _1 (p-2) \left(\delta +\lambda _1\right)+2\right)+2 V_0 \left(-\lambda _2^2+\lambda _1 \lambda _2+\delta  (p-1) \left(\delta +\lambda _1\right)-2\right)\right)}{\left(k_1^2 (p-2)-2 V_0\right) \left(2 \lambda _2^2+(p-2) (p-1) \left(\delta +\lambda _1\right){}^2+2 \lambda _2 \left(\delta  (p-1)+\lambda _1 (p-2)\right)+2\right)} \,,
\end{split} \nonumber \\
\begin{split}
&L_{r}^{2}=-\frac{2 \left(\lambda _2^2+(p-2) (p-1) \left(\delta +\lambda _1\right){}^2+\lambda _2 \left(\delta  (p-1)+\lambda _1 (p-2)\right)+2\right) }{\lambda _1^2 \left(k_1^2 (p-2)-2 V_0\right) \left(\lambda _2+\delta  (p-1)+\lambda _1 (p-2)\right){}^2} \\
& \quad \quad \times \left(2 \lambda _2^2+(p-2) (p-1) \left(\delta +\lambda _1\right){}^2+2 \lambda _2 \left(\delta  (p-1)+\lambda _1 (p-2)\right)+2\right) \,,
\end{split}   \label{Lr11} \\
&k_{2}^{2} = k_1^2-\frac{\left(\lambda _1-\lambda _2\right) \left(k_1^2 (p-2)-2 V_0\right) \left(\lambda _2+\delta  (p-1)+\lambda _1 (p-2)\right)}{\lambda _2^2+(p-2) (p-1) \left(\delta +\lambda _1\right){}^2+\lambda _2 \left(\delta  (p-1)+\lambda _1 (p-2)\right)+2} \nonumber \,.
\end{align}

The action-parameters ($\delta, \lambda_1, \lambda_2, \gamma$) may be written in terms of solution-parameters ($z, \theta, \yj$):
\begin{equation} \label{actionmetric11}
\begin{split}
 \delta=\frac{2\theta}{(p-1)\dja}\,,  \quad \lambda_{1} =\frac{-2}{\dja}\,, \quad \lambda_{2}=\frac{-2\yj}{\dja}\,, \quad  \gamma=  \frac{- 2\zeta + \frac{2}{p-1}\theta }{\dja}\,,
  \end{split}
\end{equation}
where
\begin{equation} \label{kappa1}
\begin{split}
 \dja^{2}&=2\left(\frac{\theta^{2}}{p-1}-z\zeta +2 -p -\yj^{2} \right) \\
 &=  \frac{2(p-1-\theta)\left(1+p(z-1)-z-\theta\right)}{p-1} - 2 (\yj-1) (\yj -z+1) \,,
 \end{split}
\end{equation}
where $\zeta$ can be replaced by \eqref{con2}.
Using the relations \eqref{actionmetric11}, the equations \eqref{Lr11} can be simplified as
\begin{align} 
& L_{r}^2 =  \frac{2(p-2+z-\theta)(p-2+z-\theta+\yj)}{2V_{0} -(p-2){k}_{1}^{2} } \,,  \label{Lr111} \\
& a_{0}^2 =   {\frac{2(2V_{0}(1-z)+(z(p-1)-\theta){k}_{1}^{2})}{(2-p-z+\theta-\yj)(2V_{0}-(p-2){k}_{1}^{2})}} \,,  \label{a0111}\\
& {k}_{2}^{2} = {k}_{1}^{2}+(\yj-1)\dfrac{{k}_{1}^{2}(p-2)-2V_{0}}{p-2+z-\theta} \,, \label{k2111}
\end{align}
where we choose $k_1$ as a free parameter.

\begin{table}[]
\centering
\begin{footnotesize}
\begin{tabular}{|lll|}
\hline
&  $\qquad \qquad  \qquad \qquad \quad \yj \leq 0$  &   \\
\hline
$  \qquad \quad \quad \ z<\yj $ &  $\quad \theta >p-1$   &$\zeta = 2-p+\theta -\yj$ \\
$ \quad  \quad \ \   1< z  $& $\quad  \theta <\yj  (p-1)  $  \quad &$\zeta = 2-p+\theta -\yj$  \\
\hline
 & \ {$ \qquad \qquad \qquad   \ \ 0<\yj<1$}  &  \\ \hline
 \quad \quad \quad \quad \ $ z<0$  & $ \quad \theta >p-1  $ & $\zeta = 2-p+\theta -\yj$\\
\quad \quad \ \ $ 1<z \leq 1+\yj $ & $\quad  \theta<\dfrac{p-1}{2}\left(z -\sqrt{(z-2)^{2}+\frac{4(\yj-1) (-z+1+\yj)}{p-1}}\right) $  & $\zeta = 2-p+\theta -\yj$  \\
\quad $1+\yj<z $  & $\quad \theta <\yj(p-1)  $  & $\zeta = 2-p+\theta -\yj$  \\ \hline
 &  $\qquad  \qquad  \qquad \quad 1 \leq \yj $ &  \\ \hline
$ \qquad \quad \quad \ z<0  $ & $ \quad  \theta >\yj(p-1)  $ &$\zeta = 2-p+\theta -\yj$  \\
\quad \quad  \ $  \yj <z \leq 1+\yj $ & $\quad  \theta<\dfrac{p-1}{2}\left(z -\sqrt{(z-2)^{2}+\frac{4(\yj-1) (-z+1+\yj)}{p-1}}\right)$  & $\zeta = 2-p+\theta -\yj$ \\
\ \ $ 1+\yj<z $  & $ \quad \theta <p-1  $  & $\zeta = 2-p+\theta -\yj$  \\  \hline\hline
For all cases: &   & $ k_{1}^{2}<\frac{2V_{0}(z-1)}{(p-1)z-\theta}$ \quad \quad \\
\hline
\end{tabular}
\end{footnotesize}
    \caption{Parameter range: Class I, $V_0 >0$} \label{tab:1}
\end{table}
\begin{figure}[]
 \centering
     \subfigure[$\yj=-1$]
     {\includegraphics[width=4.5cm]{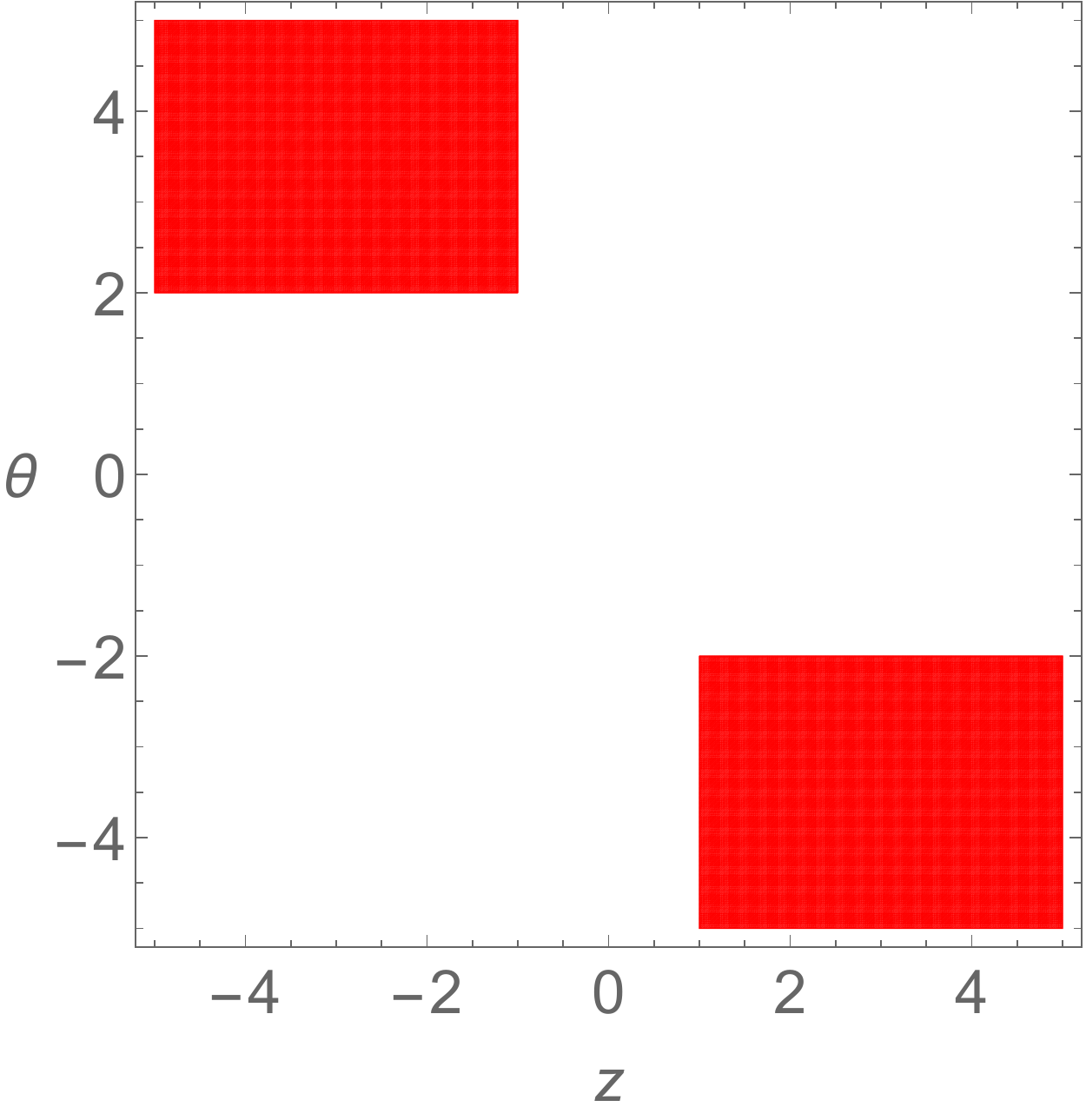} \label{}}
 \subfigure[$\yj=0$]
     {\includegraphics[width=4.5cm]{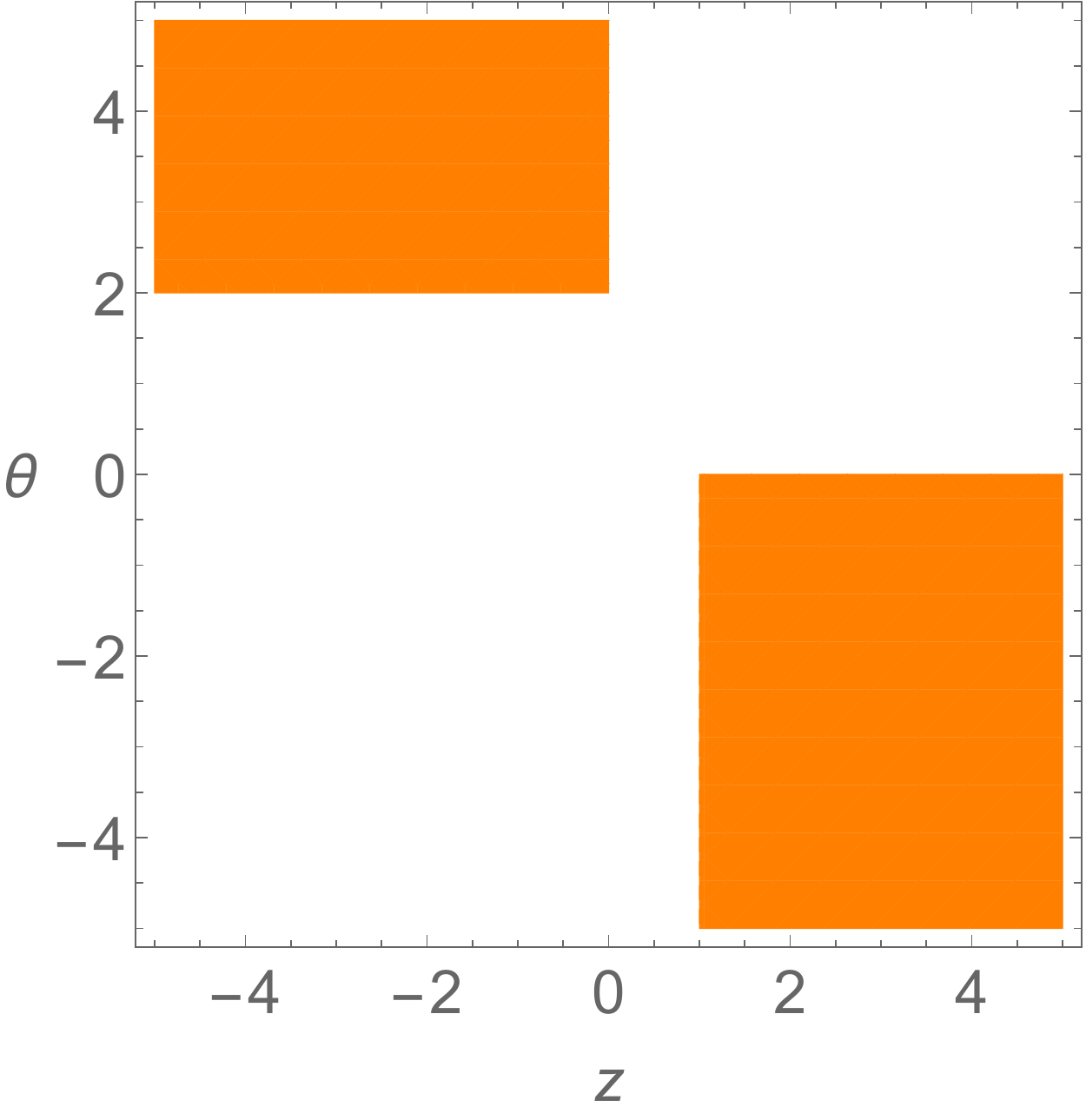} \label{}}
 \subfigure[$\yj=\frac{1}{2}$]
     {\includegraphics[width=4.5cm]{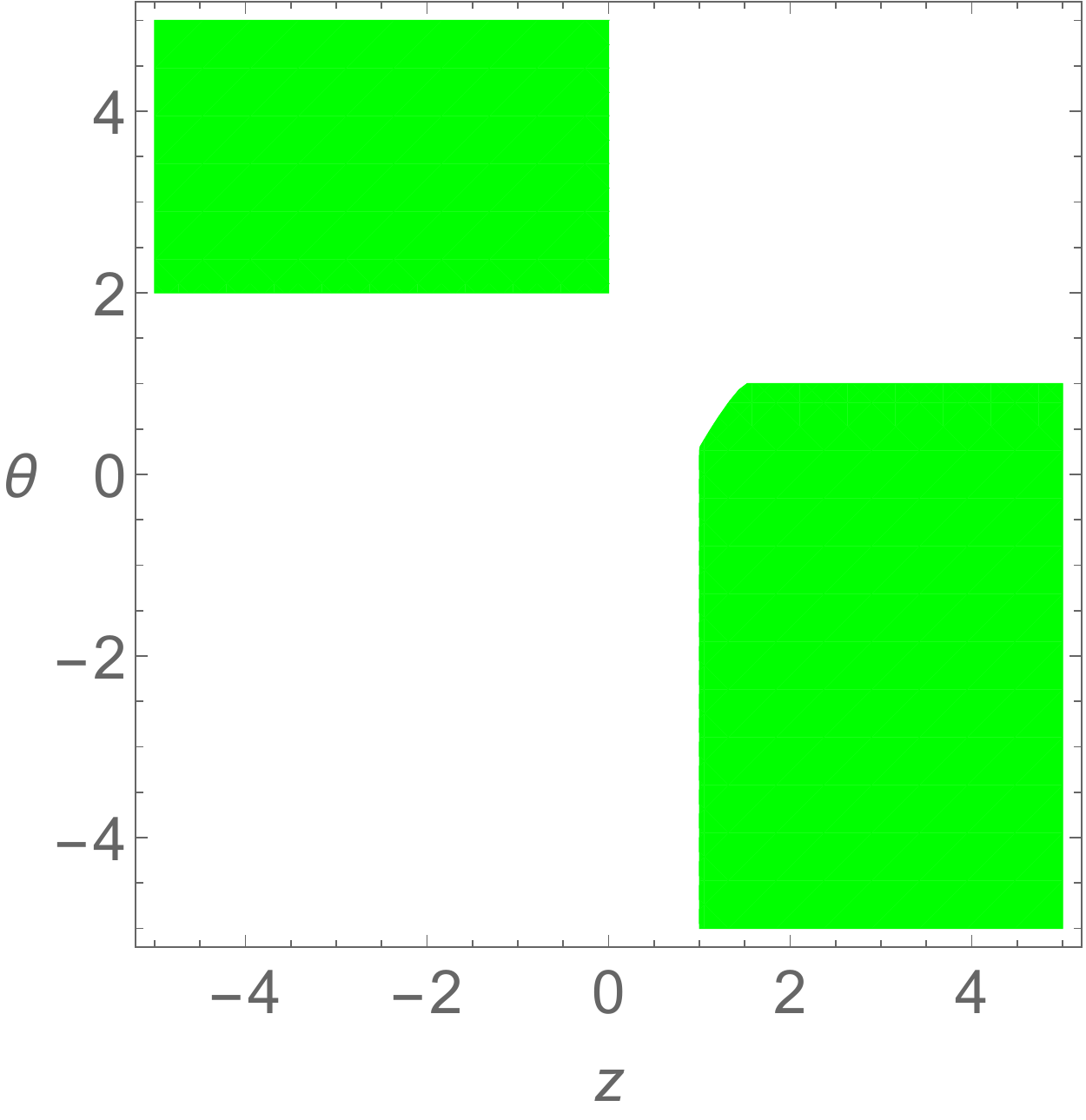} \label{}}
 \subfigure[$\yj=1$]
     {\includegraphics[width=4.5cm]{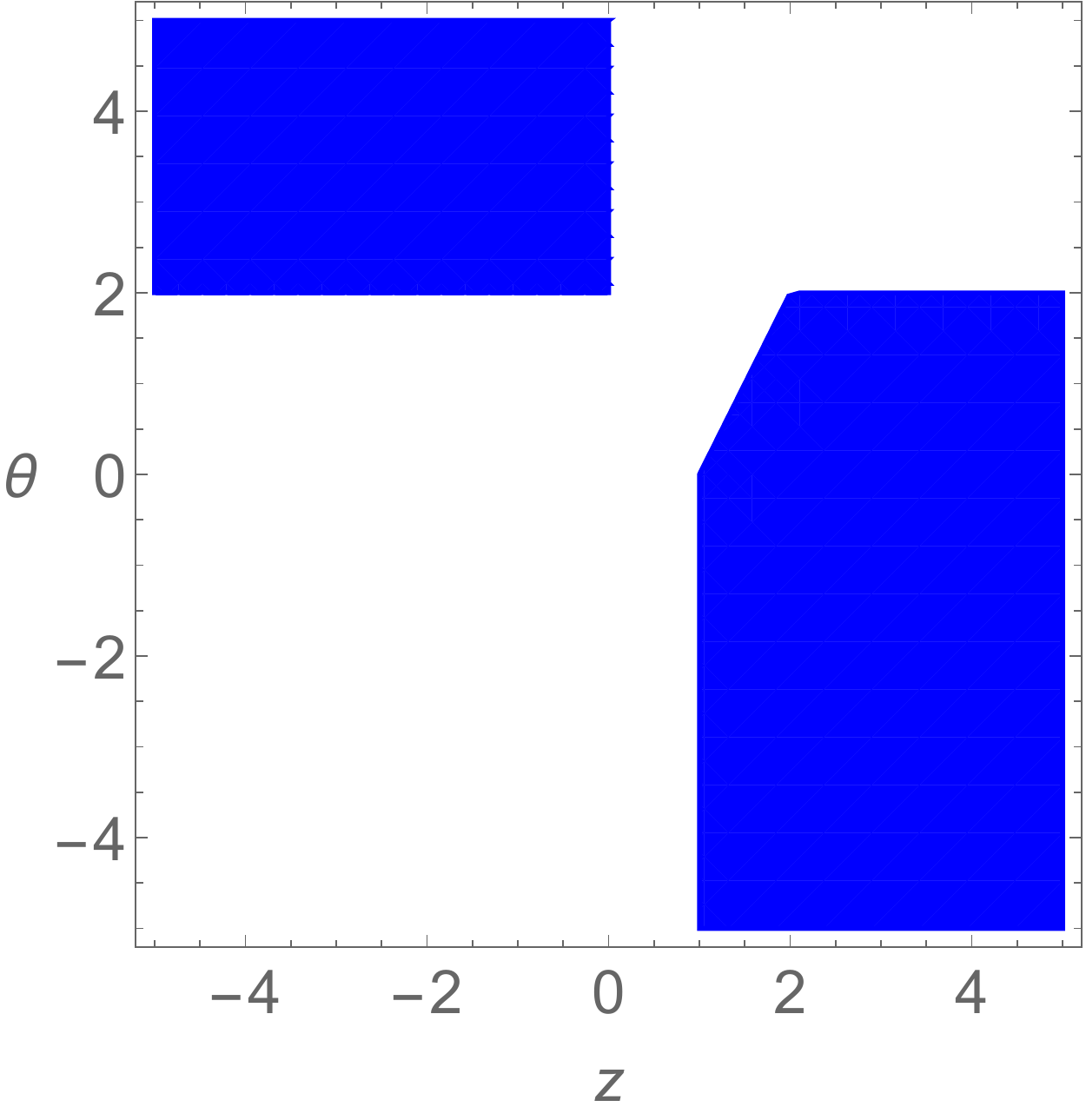} \label{}}
 \subfigure[$\yj=2$]
     {\includegraphics[width=4.5cm]{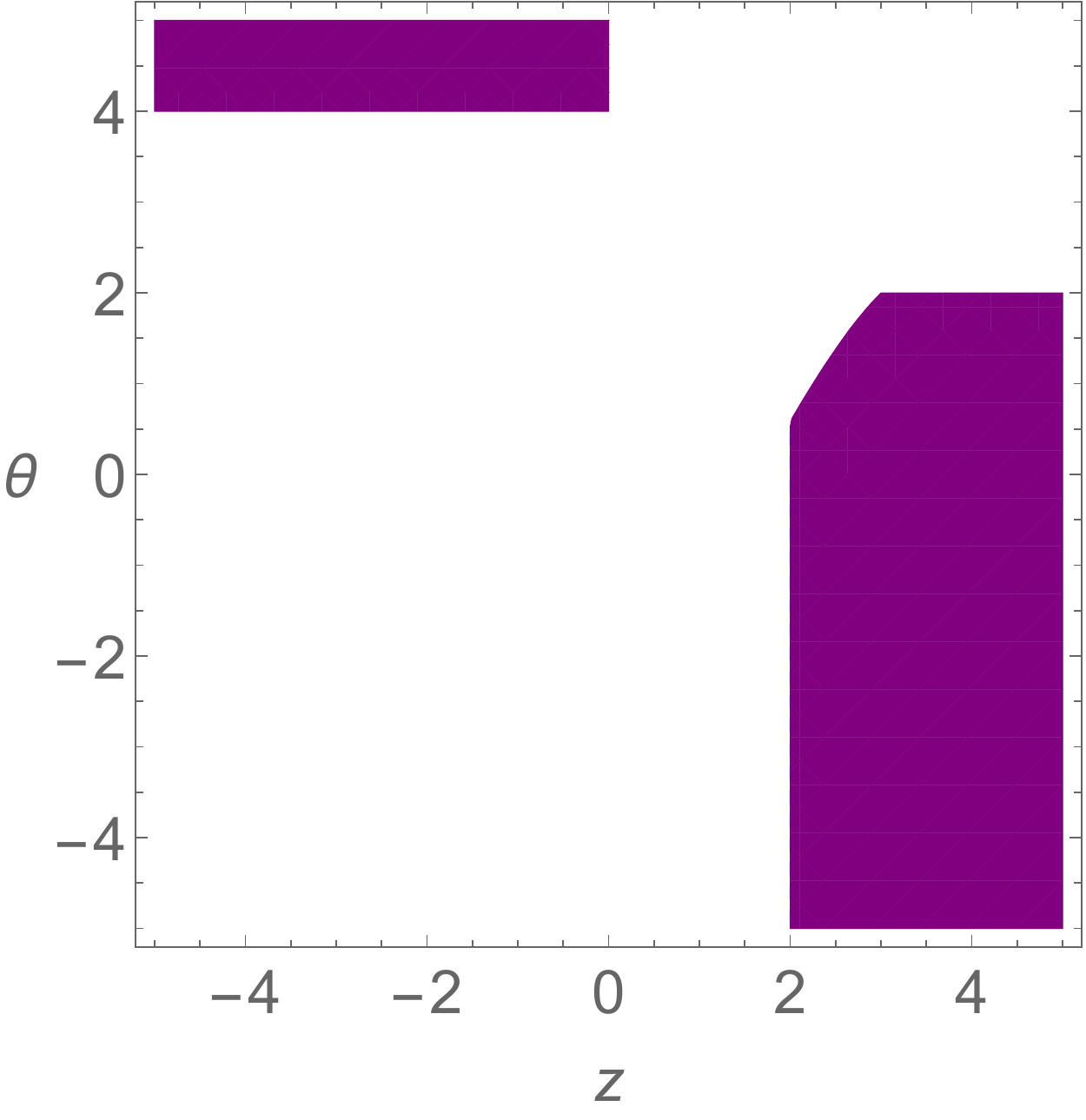} \label{}}
 \subfigure[The boundaries of (a)-(e)]
     {\includegraphics[width=4.5cm]{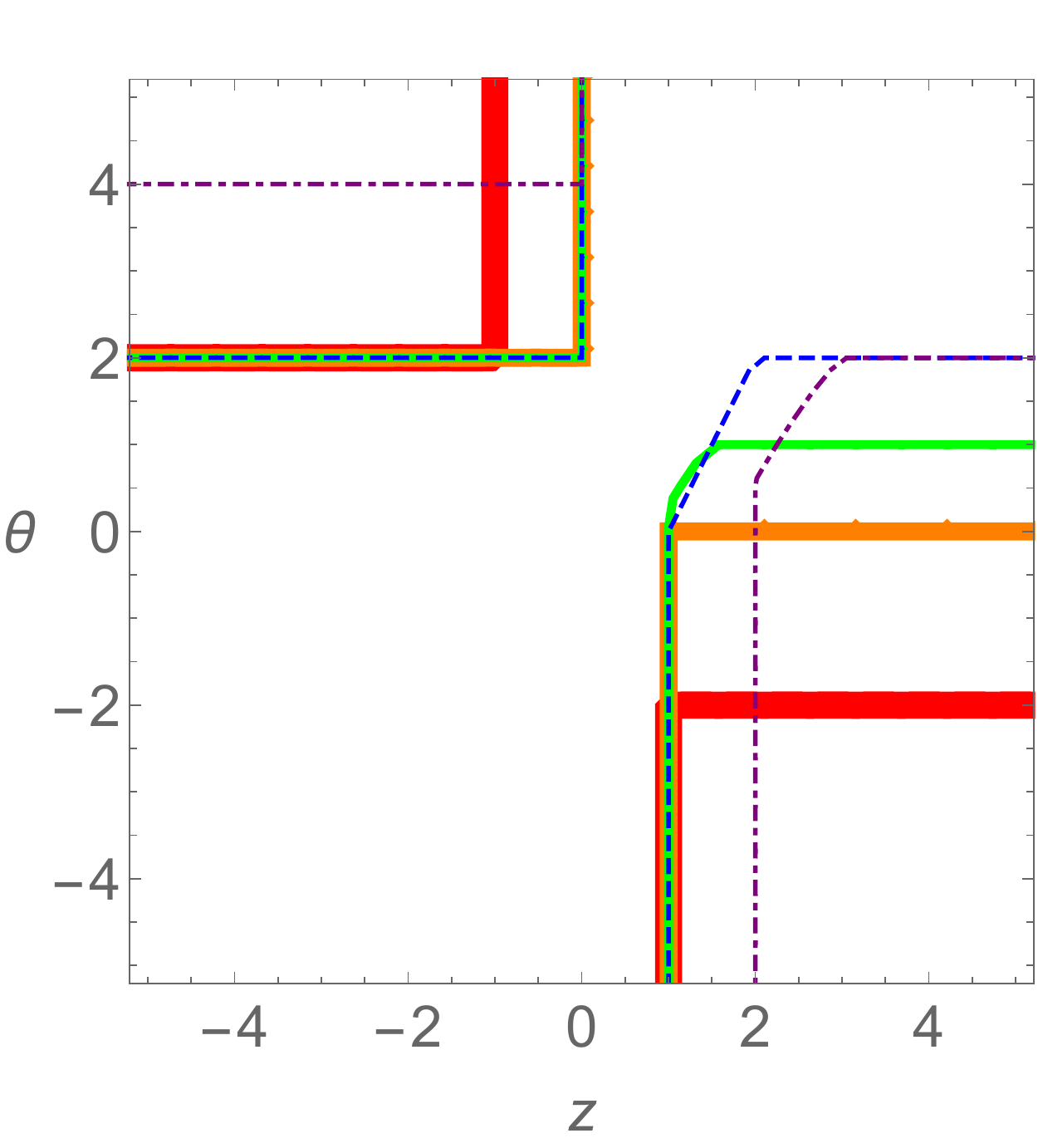} \label{}}
          \caption{Prototypical cases of Table \ref{tab:1} ($p=3$)} \label{fig:1}
\end{figure}

For given action parameters ($\delta, \lambda_1, \lambda_2, \gamma(\delta, \lambda_1,\lambda_2), V_0$) and the solution parameter $k_1$, all the solution parameters ($z, \theta, \yj, \zeta(\theta, \yj), \dja, L_r, a_0,k_2$) are fixed. Thus the total number of free parameters in the solution are four, which may be taken as ($z,\theta, \yj, k_1$).
Considering all the conditions \eqref{con111}-\eqref{con7}, we classify the allowed parameter space in Table \ref{tab:1}. For given $z$ and $\theta$, $k_1$ should be chosen to satisfy the inequality in the last column.  To get some intuition on the content of Table \ref{tab:1} it is useful to make figures representing the typical parameter ranges. Fig. \ref{fig:1} shows five prototypical cases: $\yj=-1,0,0.5,1,2$ for $p=3$. Let us start with the case $\yj=1$ (Fig. \ref{fig:1}(d)), where there are two regions: a rectangle ($z<0$)  and a pentagon ($z>1$) of which upper left corner is a line not a curve.  As $\yj$ increases (Fig. \ref{fig:1}(e)) the pentagon moves to the right ($z>\yj$) and its upper left corner becomes a curve while the rectangle moves to the up  ($\theta>2\yj$). As $\yj$ decreases (Fig. \ref{fig:1}(c,b)) the pentagon, of which upper left corner becomes curve, goes down and the rectangle does not move.  After the pentagon becomes a rectangle at $\yj=0$ it keeps going down while the rectangle in $z<0$ starts moving to the left ($z<\yj$).  For comparison we collect the boundaries of five cases in Fig. \ref{fig:1} (f).


\subsubsection{Class II: irrelevant current}

The irrelevant current means $a_0=0$ in the leading order so we start with an ansatz 
\begin{align}
\begin{split}
& \dd s^2=r^{\frac{2\theta}{p-1}}\left[-\frac{\dd t^2}{r^{2z}}+\frac{L_r^{2}\dd r^2
}{r^2}+\frac{\sum\limits_{i=1}^{p-2} \dd x_{i}^2}{r^2}+\frac{\dd y^2}{r^{2\yj}}\right]\,,   \\
& {\varphi}={\dja}\log r \,,  \quad  A=0 \,,  \quad \chi_{i}=k_{1} x_{i} \,, \quad \chi_{p-1} =k_{2} y \,,
  \end{split}
\end{align}
where $\dja, k_1$ and $k_2$ are nonzero. 

This case corresponds to $a_0=0$ in class I so $\zeta$ does not appear in the leading order solution but will be introduced when we consider a subleading order.
First, the `exponent' solution-parameters ($z, \yj, \theta$) may be expressed in terms of action-parameters ($\delta,  \lambda_1, \lambda_2 $) as
\begin{align}
\begin{split}
 z&=\frac{2-(p-1)\delta^{2}+(p-2)\lambda^{2}_{1}+\lambda^{2}_{2}}{\lambda_{1}((p-1)\delta+(p-2)\lambda_{1}+\lambda_{2})} \,,   \\
 \theta &= - \frac{(p-1)\delta}{\lambda_{1}} \,, \quad \yj =\frac{\lambda_{2}}{\lambda_{1}}\,,
 \end{split}
\end{align}
which is the same as the class I \eqref{class11} except that $\zeta$ is undetermined. It is
related to the non-existence of the constraints \eqref{con1} and \eqref{con2} in class II. Consequently, the action parameter $\gamma$ is free and ($\delta, \lambda_1, \lambda_2$) may be written in terms of three solution-parameters ($z, \theta, \yj$):
\begin{align}
\begin{split}
\delta=\frac{2\theta}{(p-1)\dja}\,,  \quad \lambda_{1} =\frac{-2}{\dja}\,, \quad \lambda_{2}=\frac{-2\yj}{\dja}\,,
  \end{split}
\end{align}
where
\begin{equation}
 \dja^{2}=  \frac{2(p-1-\theta)\left(1+p(z-1)-z-\theta\right)}{p-1} - 2 (\yj-1) (\yj -z+1) \,.
\end{equation}
The coefficient parameters read
\begin{equation} \label{jjj11}
\begin{split}
&\dja= -\frac{2}{\lambda_1}  \,, \\
&L_{r}^{2}=\, \frac{(2-p+\theta-\yj-z)\left(\theta-(p-1)z\right)}{V_{0}} \,, \\
&k_{1}^{2} = \dfrac{2V_{0}(1-z)}{\theta-(p-1)z} \,, \\
&k_{2}^{2} = \dfrac{2V_{0}(\yj-z)}{\theta-(p-1)z} \,.
\end{split}
\end{equation}

After turning on the subleading gauge field mode generating a constant electric flux proportional to $a_0$
\begin{equation} \label{class2ar}
A_{t}(r) = \, a_0 r^{ \zeta-z} \,,
\end{equation}
we find
\begin{equation} \label{zeta1}
\begin{split}
\zeta& = p-2-\frac{p-3}{p-1}\theta - \gamma \dja + \yj \,,  \\
\dja^2  &= { \frac{2(p-1-\theta)\left(1+p(z-1)-z-\theta\right)}{p-1} - 2( \yj-1) (\yj -z+1)} \,,
\end{split}
\end{equation}
where $\zeta$ is a function of a free action parameter $\gamma$.
This gauge field mode backreacts on metric and $\varphi$ at quadratic order as
\begin{equation} \label{class2ar1}
\sim r^\beta\,,  \qquad \mathrm{where} \quad  \beta := p-2+\zeta-\theta+\yj\,,
\end{equation}
which gives a constraint on $\zeta$ because $\beta$ should be positive(negative) if the IR is at $r \rightarrow 0(\infty)$. This constraint (inequality) was summarized in the third column in Table \ref{tab:2}. After considering all conditions  \eqref{con111}-\eqref{con7}, we find that the parameter space of $z,\theta,\yj$ are the same as class I as shown in Table \ref{tab:2}.  Therefore, Fig \ref{fig:1} are valid also for class 2.
For given $z$ and $\theta$, $\gamma$ should be chosen to satisfy the inequality in the last column because $\zeta$ is a function of $\gamma$ for given $z,\theta$ and $\yj$ \eqref{zeta1}.

\begin{table}
\centering
\begin{tabular}{|lll|}
\hline
 &  $\qquad \qquad  \qquad \qquad \quad \yj \leq 0$  &   \\
\hline
$  \qquad \quad \  z<\yj $ &  $\quad \theta >p-1  $ &$ \quad \zeta > 2-p+\theta-\yj$ \\
$ \quad  \  \ 1< z  $& $\quad  \theta <\yj (p-1) $& $ \quad \zeta < 2-p+\theta-\yj $\\
\hline
 & \ {$ \qquad \qquad \qquad \ \   0<\yj<1$}  & \\ \hline
 $\qquad \ \ \ \ \ z<0 $  & $ \quad \theta >p-1  $ & $ \quad \zeta > 2-p+\theta-\yj $ \\
$\quad \ \ 1<z \leq 1+\yj $ & $\quad  \theta<\dfrac{p-1}{2}\left(z -\sqrt{(z-2)^{2}+\frac{4(\yj-1) (-z+1+\yj)}{p-1}}\right) $ & $  \quad   \zeta < 2-p+\theta-\yj $  \\
$1+\yj<z $  & $\quad \theta <\yj(p-1)  $ & $ \quad  \zeta < 2-p+\theta-\yj $  \\ \hline
 &  $\qquad  \qquad  \qquad \quad 1 \leq \yj $ &  \\ \hline
$ \qquad \ \ \ \  z<0  $ & $ \quad  \theta >\yj(p-1)  $ &  $ \quad  \zeta > 2-p+\theta-\yj $  \\
\quad \ \ $  \yj <z \leq 1+\yj $ & $\quad  \theta<\dfrac{p-1}{2}\left(z -\sqrt{(z-2)^{2}+\frac{4(\yj-1) (-z+1+\yj)}{p-1}}\right)$ & $ \quad  \zeta < 2-p+\theta-\yj $ \\
$ 1+\yj<z $  & $ \quad \theta <p-1  $ & $  \quad \zeta < 2-p+\theta-\yj $ \\
\hline
\end{tabular}
\caption{Parameter range: Class II, $V_0 >0$} \label{tab:2}
\end{table}

\subsection{Irrelevant axion}  \label{sec:22}

Irrelevant axion means that $k_1=k_2=0$($\chi_i = \chi_{p-1} = 0$) at  leading order in the IR. In principle,
there may be anisotropic solutions generated by the subleading axion mode due to anisotropy in the action, $\lambda_1 \ne \lambda_2$. However, after turning on the subleading axion mode ($\chi_i = k_1 x, \, \chi_{p-1} = k_2 y$)  we find that $\lambda_1$ must be the same as $\lambda_2$ to satisfy the equations of motion in the subleading order. 
This does not mean that this case becomes the isotropic case. Because $k_1 \ne k_2$ in the sub-leading order, it is a new kind of anisotropic solution with $\lambda_1=\lambda_2$. 

\subsubsection{Class III: marginally relevant current}

The irrelevant axion means $k_1=k_2=0$ in the leading order so we start with an ansatz 
\begin{align} \label{class3sol1}
\begin{split}
& \dd s^2=r^{\frac{2\theta}{p-1}}\left[-\frac{\dd t^2}{r^{2z}}+\frac{L_r^{2}\dd r^2
}{r^2}+\frac{\sum\limits_{i=1}^{p-2} \dd x_{i}^2}{r^2}+\frac{\dd y^2}{r^{2\yj}}\right]\,,   \\
& {\varphi}={\dja}\log r \,,  \quad A= a_0 \,r^{\zeta-z}\,\dd t \,,    \quad \chi_{i}=0 \,, \quad \chi_{p-1} =0 \,,
  \end{split}
\end{align}
where $ \dja$ and $a_0$ are nonzero. 

By the equations of motion, solution-parameters ($z, \theta, \yj, \dja, \zeta$) may be expressed in terms of two action-parameters ($\delta,  \gamma $) as
\begin{align} \label{kappa3sol2}
\begin{split}
 z&=\frac{\dja ^2 \left(\delta ^2 (p-1)-2\right)-4 (p-1)}{2 (p-1) (\delta  \dja -2)}\,,  \quad   \dja = \frac{2 (p-1)}{\gamma +\delta  (p-2)}  \,,   \\
 \theta &=\frac{(p-1)\delta \dja}{2}  \,, \quad  \yj=1 \,, \quad \zeta =  \frac{(\delta-\gamma)\dja}{2} \,.
 \end{split}
\end{align}
They satisfy the following constraint 
\begin{equation} \label{con23}
\zeta  = -(p-1 -\theta) \equiv - d_\theta \equiv \zeta_I \,,
\end{equation}
which corresponds to the relation between action parameters
\begin{equation} \label{con13}
 \gamma =  (2-p)\delta+\frac{2(p-1)}{\dja} \,.
\end{equation}
%
The remaining solutions parameters are
\begin{align} 
L_{r}^{2}&=\frac{(z+p-\theta -1)(z+p-\theta -2)}{V_{0}} \,, \label{Lr033}\\
a_0^2 &= {\frac{2(-1+z)}{-1+p+z-\theta}} \,, \label{a03}
\end{align}
where $\zeta \ne z$ and $\zeta \ne z-1$ are assumed.%

The action-parameters ($\delta, \gamma$) may be written in terms of two solution-parameters ($z, \theta$):
\begin{align} \label{actionmetric2}
\begin{split}
 \delta=\frac{2\theta}{(p-1)\dja}\,, \quad \gamma=  \frac{- 2\zeta + \frac{2}{p-1}\theta }{\dja}\,,
  \end{split}
\end{align}
where
\begin{equation} \label{kappa3}
\begin{split}
 \dja^{2}&=2\left(\frac{\theta^{2}}{p-1}-z\zeta +1-p \right) \\
 &=  \frac{2(p-1-\theta)\left(1+p(z-1)-z-\theta\right)}{p-1}  \,.
 \end{split}
\end{equation}

Note that all formulas so far are independent of $\lambda_1$ and $\lambda_2$ and they are free. However, after introducing the subleading axion mode ($\chi_i = k_1 x, \chi_{p-1} = k_2 y$) we find $\lambda_1 = \lambda_2 :=\lambda$ and they backreact on metric and $\varphi$ at quadratic order as
\begin{equation} \label{class3mode}
\sim r^\beta\,,  \qquad \mathrm{where} \quad  \beta := 2+\dja \lambda,
\end{equation}
which gives a constraint on $\dja\lambda$ because $\beta$ should be positive(negative) if the IR is at $r \rightarrow 0(\infty)$. This constraint (inequality) was summarized in the third column in \eqref{con33} below. After considering all conditions  \eqref{con111}-\eqref{con7}, we find that the parameter space of $z,\theta$ are
\begin{align} \label{con33}
\begin{split}
&\qquad z < 0 \,,    \ \quad  \theta > p-1 \,, \quad \qquad \qquad \, \dja \lambda > - 2 \,, \\
& \,1<z\le 2 \,, \ \ \,  \ \theta < (z-1)(p-1) \,, \quad \  \dja \lambda < - 2  \,\,, \\
&\,2<z \,, \quad \qquad \theta <p-1 \,, \quad \qquad  \qquad \,  \dja \lambda < -2 \,,
\end{split}
\end{align}
which is the same as the case I for $\yj=1$ and represented in Fig \ref{fig:1}(d).
$\lambda$ should be chosen for given $z, \theta$ to satisfy the last inequality in \eqref{con33} with \eqref{kappa3}. If $\lambda$ is given, $\dja$ should be chosen to satisfy the last inequality in  \eqref{con33}, which further restricts the range of $z$ and $\theta$.

All formulas in this section are consistent with the formulas in Class I with replacements: $\yj=1$, $\lambda_1=\lambda_2$ and $k_1=k_2=0$.

\subsubsection{Class IV: irrelevant current}

This class correspond to $k_1=k_2=0=a_0$ in the leading order so we start with an ansatz 
\begin{align} \label{class4sol1}
\begin{split}
& \dd s^2=r^{\frac{2\theta}{p-1}}\left[-\frac{\dd t^2}{r^{2z}}+\frac{L_r^{2}\dd r^2
}{r^2}+\frac{\sum\limits_{i=1}^{p-2} \dd x_{i}^2}{r^2}+\frac{\dd y^2}{r^{2\yj}}\right]\,,   \\
& {\varphi}={\dja}\log r \,,  \quad A=  0\,,    \quad \chi_{i}=0 \,, \quad \chi_{p-1} =0 \,,
  \end{split}
\end{align}
where the solution variables are determined by the action variable $(\delta, V_0)$ as follows:
\begin{align}
&z=1\,, \qquad \yj =1 \,, \qquad \dja =   \frac{2 \delta  (p-1)}{\delta ^2 (p-1)-2}  \,,   \\
& \theta =\frac{(p-1)\delta \dja}{2}=\frac{\delta ^2 (p-1)^2}{\delta ^2 (p-1)-2}  \,, \qquad
L_{r}^{2}=\frac{(p-\theta )(p-1-\theta) }{V_0} \,.
\end{align}
We may deduce $z=1$ from \eqref{a03} by setting  $a_0 = 0$. 
The relation between $\dja$ and $\delta$ may be understood by requiring $z=1$ in the first equation of \eqref{kappa3sol2}.
$L_r$ is can be read from \eqref{Lr033} with $z=1$.
The action variable $\delta$ reads 
\begin{equation}
 \delta=\frac{2\theta}{(p-1)\dja} \qquad \mathrm{with} \qquad
 \dja^{2}=\frac{2\theta(1-p+\theta)}{p-1} \,.
\end{equation}
in terms of solution variables. 

Note that all formulas so far are independent of $\lambda_1, \lambda_2$ and $\gamma$ because both axions and current are irrelevant. By turning on the subleading gauge field mode
\begin{equation}
A_{t}(r) = \, a_0 r^{ \zeta-1}  \,,
\end{equation}
we find
\begin{equation} \label{zeta14}
\zeta = p-1-\frac{p-3}{p-1}\theta - \gamma \dja  \,,
\end{equation}
where $\zeta$ is a function of a free action parameter $\gamma$.
This gauge field mode backreact on metric and $\varphi$ at quadratic order as
\begin{equation} \label{final1444}
\sim r^\beta\,,  \qquad \mathrm{where} \quad  \beta := p-1+\zeta-\theta \,.
\end{equation}
This mode analysis is parallel to class II from \eqref{class2ar} to \eqref{class2ar1}.
If we turn on the subleading axion mode ($\chi_i = k_1 x, \chi_{p-1} = k_2 y$) we find $\lambda_1 = \lambda_2 :=\lambda$ and they backreact on metric and $\varphi$ at quadratic order as
\begin{equation} \label{final2444}
\sim r^\beta\,,  \qquad \mathrm{where} \quad  \beta := 2+\dja \lambda \,.
\end{equation}
This mode analysis is parallel to \eqref{class3mode} in class III.

After considering all conditions we listed in case I with \eqref{final1444} and \eqref{final2444}, we find that the parameter space is
\begin{equation} \label{con44}
z =1 \,,    \ \quad  \theta <  0 \,, \quad  \zeta < \theta+1-p \,, \quad \dja \lambda < - 2  \,.
\end{equation}

\subsection{Marginally relevant and irrelevant axion}

In this subsection we consider the case that only one of the $k_i$ is nonzero. 
This is a hybrid of the class I and III ($a_0 \ne 0$) ; and the class II and IV ($a_0 =0$). 
The class I-i and II-i means $k_1$ is nonzero and the class I-ii and II-ii means $k_2$ is nonzero.

\subsubsection{Class I-i: marginally relevant current}

We assume that the classical solutions are written as
\begin{equation} \label{class1sol}
\begin{split}
& \dd s^2=r^{\frac{2\theta}{p-1}}\left[-\frac{\dd t^2}{r^{2z}}+\frac{L_r^{2}\dd r^2
}{r^2}+\frac{\sum\limits_{i=1}^{p-2} \dd x_{i}^2}{r^2}+\frac{\dd y^2}{r^{2\yj}}\right]\,,   \\
& {\varphi}={\dja}\log r \  \,,  \quad  A=a_{0}\,r^{\zeta-z}\,\dd t \,,  \quad \chi_{i}=k_{1} x_{i} \,, \quad \chi_{p-1} =0\,.
  \end{split}
\end{equation}
By the equations of motion,  the `exponent' solution-parameters ($z, \yj, \theta,  \zeta$) may be expressed in terms of  action-parameters ($\delta,  \lambda_1 , \gamma $) as
    \begin{equation}
        \begin{split}
            z &= -\frac{2+\gamma ^2+((p-5) p+5)\delta ^2+(p-2) \left(2 \left(\delta +\lambda _1\right)\gamma+\lambda _1 \left(2(p-2) \delta +(p-1)\lambda _1 \right)\right)}{(\gamma -\delta )\lambda _1}\,,\\
            \theta &= \frac{(1-p)\delta}{\lambda _1} , \qquad \yj =-\frac{\gamma +(p-2) \left(\delta +\lambda _1\right)}{\lambda _1} , \qquad \zeta = \frac{\gamma - \delta}{\lambda_{1}}   \nonumber \,.
        \end{split}
    \end{equation}
Note that $\lambda_2$ does not contribute to the IR solution because $k_2=0$. A quick way to see this solutions is to solve \eqref{con1} for $\lambda_2$ and plugging it to \eqref{class11} making it $\lambda_2$ independent. 
They are not all independent and there is a constraint between  solution-parameters ($\theta, \zeta, \yj$)
\begin{equation} \label{con277}
\zeta  = -(p-2 -\theta) -\yj \equiv - (d_\theta-1) -\yj \equiv \zeta_I \,,
\end{equation}
which does not give any relation between action-parameters contrary to \eqref{con1}.
The `coefficient' parameters are solved as
%
    \begin{align}
        & \dja = -\frac{2}{\lambda_1} \,,  \\
        & L_r^{2} =\frac{(p-2+z-\theta +\xi) (z-\theta +(p-2)\xi)}{V_0} \,, \\
        &a_{0}^2 = \frac{2 (z-\xi )}{p-2+z-\theta +\xi } \,, \\
        &k_{1}^{2} = \frac{2V_0 (\xi -1)}{z-\theta +(p-2)\xi} \,. 
    \end{align}
It can be understood by \eqref{Lr111}-\eqref{k2111} by solving for $k_1$ with $k_2=0$ and plugging it back. 

The action-parameters ($\delta, \lambda_1,  \gamma$) may be written in terms of solution-parameters ($z, \theta, \yj$):
\begin{equation} \label{actionmetric1}
\begin{split}
 \delta=\frac{2\theta}{(p-1)\dja}\,,  \quad \lambda_{1} =\frac{-2}{\dja}\,, \quad   \gamma=  \frac{- 2\zeta + \frac{2}{p-1}\theta }{\dja}\,,
  \end{split}
\end{equation}
where
\begin{equation} \label{kappa1}
\begin{split}
 \dja^{2}&=2\left(\frac{\theta^{2}}{p-1}-z\zeta +2 -p -\yj^{2} \right) \\
 &=  \frac{2(p-1-\theta)\left(1+p(z-1)-z-\theta\right)}{p-1} - 2 (\yj-1) (\yj -z+1) \,,
 \end{split}
\end{equation}
where $\zeta$ can be replaced by \eqref{con277}.

Here, all formulas are independent of $\lambda_2$. By considering the sub-leading axion mode $k_2 y$ we find that it backreacts on metric and $\varphi$ at quadratic order as
    \begin{equation}
        \sim r^\beta\,,  \qquad \mathrm{where} \quad  \beta := \dja \lambda_{2} + 2\yj \,,
    \end{equation}
which gives a constraint on $\dja\lambda_2$ because $\beta$ should be positive(negative) if the IR is at $r \rightarrow 0(\infty)$. This and all other conditions  \eqref{con111}-\eqref{con7} give us the parameter space shown in Table \ref{tab:3}.
\begin{table}
\centering
\begin{footnotesize}
\begin{tabular}{|lll|}
\hline
&  $\qquad \qquad  \qquad \qquad \quad \yj \leq 0$  &   \\
\hline
$  \qquad \quad \quad \ z<\yj $ &  $\quad \theta >p-1$   &$\zeta = 2-p+\theta -\yj, \quad \frac{\lambda_{2}}{\lambda_{1}}<\yj$ \\
\hline
 & \ {$ \qquad \qquad \qquad \ \  0<\yj<1$}  &  \\ \hline
 \quad \quad \quad \quad \ $ z<0$  & $ \quad \theta >p-1  $ & $\zeta = 2-p+\theta -\yj, \quad \frac{\lambda_{2}}{\lambda_{1}}<\yj$\\
\quad $1+\yj<z $  & $\quad \theta <\yj(p-1)  $  & $\zeta = 2-p+\theta -\yj, \quad \frac{\lambda_{2}}{\lambda_{1}}>\yj$  \\ \hline
 &  $\qquad  \qquad  \qquad \quad 1 \leq \yj $ &  \\ \hline
\quad \quad  \ $  \yj <z \leq 1+\yj $ & $\quad  \theta<\dfrac{p-1}{2}\left(z -\sqrt{(z-2)^{2}+\frac{4(\yj-1) (-z+1+\yj)}{p-1}}\right)$  & $\zeta = 2-p+\theta -\yj, \quad \frac{\lambda_{2}}{\lambda_{1}}>\yj$ \\
\ \ $ 1+\yj<z $  & $ \quad \theta <p-1  $  & $\zeta = 2-p+\theta -\yj, \quad \frac{\lambda_{2}}{\lambda_{1}}>\yj$  \\  \hline
\end{tabular}
\end{footnotesize}
    \caption{Parameter range: Class I-i, $V_0 >0$} \label{tab:3}
\end{table}

\subsubsection{Class I-ii: marginally relevant current}

We assume that the classical solutions are written as
\begin{equation} \label{class1sol}
\begin{split}
& \dd s^2=r^{\frac{2\theta}{p-1}}\left[-\frac{\dd t^2}{r^{2z}}+\frac{L_r^{2}\dd r^2
}{r^2}+\frac{\sum\limits_{i=1}^{p-2} \dd x_{i}^2}{r^2}+\frac{\dd y^2}{r^{2\yj}}\right]\,,   \\
& {\varphi}={\dja}\log r \  \,,  \quad  A=a_{0}\,r^{\zeta-z}\,\dd t \,,  \quad \chi_{i}=0 \,, \quad \chi_{p-1} =k_2 y\,.
  \end{split}
\end{equation}
By the equations of motion, the `exponent' solution-parameters ($z, \yj, \theta,  \zeta$) may be expressed in terms of action-parameters ($\delta,  \lambda_2 , \gamma $) as
%
    \begin{equation}
        \begin{split}
            z &= \frac{\gamma ^2+2 (p-2)+2 \gamma  \left(\lambda _2+ (p-2)\delta\right)-(p-2)\delta ^2+\lambda _2 \left(2 (p-2)\delta+(p-1)\lambda _2\right)}{(\gamma -\delta ) \left(\gamma +\lambda _2+(p-2)\delta \right)},  \\
            \theta &= \frac{(p-1)(p-2)\delta}{\gamma +\lambda _2+ (p-2)\delta}, \qquad \yj = -\frac{ (p-2)\lambda _2}{\gamma +\lambda _2+(p-2)\delta}, \qquad \zeta = -\frac{(p-2) (\gamma -\delta )}{\gamma +\lambda _2+(p-2)\delta} \,.
        \end{split}
    \end{equation}
Note that $\lambda_1$ does not contribute to the IR solution because $k_1=0$. A quick way to see this solutions is to solve \eqref{con1} for $\lambda_1$ and plugging it to \eqref{class11} making it $\lambda_1$ independent. 
They are not all independent and there is a constraint between solution-parameters ($\theta, \zeta, \yj$)
\begin{equation} \label{con288}
\zeta  = -(p-2 -\theta) -\yj \equiv - (d_\theta-1) -\yj \equiv \zeta_I \,.
\end{equation}
which does not give any relation between action-parameters contrary to \eqref{con1}.
The `coefficient' parameters are solved as
%
    \begin{align}
        & \dja = - \frac{2\yj}{\lambda_{2}} \,,  \\
        & L_r^{2} = \frac{(p-2+z-\theta) (p-2+z-\theta +\xi)}{V_0} \,, \\
        &a_{0}^2 = \frac{2 (z-1)}{p-2+z-\theta +\xi}\,, \\
        & k_{2}^{2} = \frac{2V_0 (1-\xi ) }{p-2+z-\theta} \,.
    \end{align}
It can be understood by \eqref{Lr111}-\eqref{k2111} by solving for $k_2$ with $k_1=0$ and plugging it back. 

The action-parameters ($\delta, \lambda_2,  \gamma$) may be written in terms of solution-parameters ($z, \theta, \yj$):
\begin{equation} \label{actionmetric1}
\begin{split}
 \delta=\frac{2\theta}{(p-1)\dja}\,,  \quad \lambda_{2} =\frac{-2\yj}{\dja}\,, \quad   \gamma=  \frac{- 2\zeta + \frac{2}{p-1}\theta }{\dja}\,,
  \end{split}
\end{equation}
where
\begin{equation} \label{kappa1}
\begin{split}
 \dja^{2}&=2\left(\frac{\theta^{2}}{p-1}-z\zeta +2 -p -\yj^{2} \right) \\
 &=  \frac{2(p-1-\theta)\left(1+p(z-1)-z-\theta\right)}{p-1} - 2 (\yj-1) (\yj -z+1) \,,
 \end{split}
\end{equation}
where $\zeta$ can be replaced by \eqref{con288}.

Similarly to the case I-i, all formulas here are independent of $\lambda_1$. By considering the sub-leading axion mode $k_1 x_i$ we find that it backreacts on metric and $\varphi$ at quadratic order as
    \begin{equation}
        \sim r^\beta\,,  \qquad \mathrm{where} \quad  \beta := \dja \lambda_{1} + 2 \,,
    \end{equation}
which gives a constraint on $\dja\lambda_1$ because $\beta$ should be positive(negative) if the IR is at $r \rightarrow 0(\infty)$. This and all other conditions  \eqref{con111}-\eqref{con7} give us the parameter space shown in Table \ref{tab:4}.
\begin{table}
\centering
\begin{footnotesize}
\begin{tabular}{|lll|}
\hline
&  $\qquad \qquad  \qquad \qquad \quad \yj \leq 0$  &   \\
\hline
$ \quad  \quad \ \   1< z  $& $\quad  \theta <\yj  (p-1)  $  \quad &$\zeta = 2-p+\theta -\yj, \quad \frac{\lambda_{1}}{\lambda_{2}}<\yj^{-1}$  \\
\hline
 & \ {$ \qquad \qquad \qquad \ \   0<\yj<1$}  &  \\ \hline
\quad \quad \ \ $ 1<z \leq 1+\yj $ & $\quad  \theta<\dfrac{p-1}{2}\left(z -\sqrt{(z-2)^{2}+\frac{4(\yj-1) (-z+1+\yj)}{p-1}}\right) $  & $\zeta = 2-p+\theta -\yj, \quad \frac{\lambda_{1}}{\lambda_{2}}>\yj^{-1}$  \\
\quad $1+\yj<z $  & $\quad \theta <\yj(p-1)  $  & $\zeta = 2-p+\theta -\yj, \quad \frac{\lambda_{1}}{\lambda_{2}}>\yj^{-1}$  \\ \hline
 &  $\qquad  \qquad  \qquad \quad 1 \leq \yj $ &  \\ \hline
$ \qquad \quad \quad \ z<0  $ & $ \quad  \theta >\yj(p-1)  $ &$\zeta = 2-p+\theta -\yj, \quad \frac{\lambda_{1}}{\lambda_{2}}<\yj^{-1}$  \\\hline
\end{tabular}
\end{footnotesize}
    \caption{Parameter range: Class I-ii, $V_0 >0$} \label{tab:4}
\end{table}

\subsubsection{Class II-i:  irrelevant current}

We assume that the classical solutions are written as
\begin{equation} \label{class1sol}
\begin{split}
& \dd s^2=r^{\frac{2\theta}{p-1}}\left[-\frac{\dd t^2}{r^{2z}}+\frac{L_r^{2}\dd r^2
}{r^2}+\frac{\sum\limits_{i=1}^{p-2} \dd x_{i}^2}{r^2}+\frac{\dd y^2}{r^{2\yj}}\right]\,,   \\
& {\varphi}={\dja}\log r \  \,,  \quad  A=0 \,,  \quad \chi_{i}=k_{1} x_{i} \,, \quad \chi_{p-1} =0 \,.
  \end{split}
\end{equation}
By the equations of motion, the `exponent' solution-parameters ($z, \yj, \theta$) may be expressed in terms of action-parameters ($\delta,  \lambda_1 $) as
    \begin{equation}
            z = \yj  \,, \qquad
            \theta = \frac{(1-p)\delta}{\lambda_{1}} , \qquad \yj =\frac{2-(p-1)\delta ^2+ (p-2)\lambda _1^2}{\left((p-1)\delta+(p-2)\lambda _1\right)\lambda _1 } \,.
    \end{equation}
Note that $\lambda_2$ and $\gamma$ does not contribute to the IR solution because $k_2=a_0=0$. 
The first equation can be understood by \eqref{jjj11} where we set $k_2=0$.
The `coefficient' parameters are solved as
    \begin{align}
        &  \dja = -\frac{2}{\lambda_{1}}\,,  \\
        & L_r^{2} =  \frac{(p-2-\theta+2 \yj) ((p-1) \yj-\theta )}{V_0} \,, \\
        &k_{1}^{2} = \frac{2 V_0 (\yj-1)}{(p-1)\yj-\theta}    \,. 
    \end{align}
It can be understood by \eqref{jjj11} by plugging $z=\yj$.

The action-parameters ($\delta, \lambda_1$) may be written in terms of solution-parameters ($\theta, \yj$):
\begin{equation} \label{actionmetric1}
\begin{split}
 \delta=\frac{2\theta}{(p-1)\dja}\,,  \quad \lambda_{1} =\frac{-2}{\dja}\,, 
  \end{split}
\end{equation}
where 
\begin{equation} \label{kappa1}
\begin{split}
 \dja^{2}&=  \frac{2(p-1-\theta)\left(1+p(\yj-1)-\yj-\theta\right)}{p-1} - 2 (\yj-1) \,.
 \end{split}
\end{equation}

Note that all formulas so far are independent of $\lambda_2$ and $\gamma$ because one axion ($k_2 y$) and the current are irrelevant. By turning on the subleading gauge field mode
\begin{equation}
A_{t}(r) = \, a_0 r^{ \zeta-\yj}  \,,
\end{equation}
we find
\begin{equation} \label{zeta14}
        \zeta = \frac{ ( p(p-3)+2(\theta+1) ) + (1-p)(\gamma \dja -\yj+\theta)}{p-1} \,,
\end{equation}
where $\zeta$ is a function of a free action parameter $\gamma$.
This gauge field mode backreact on metric and $\varphi$ at quadratic order as
\begin{equation} \label{final1}
        \sim r^{\beta_{1}}\,,  \qquad \mathrm{where} \quad  \beta_{1} := p-2+\zeta -\theta +\yj \,.
\end{equation}
If we turn on the subleading axion mode ($\chi_{p-1} = k_2 y$) we find
    \begin{equation}
        \lambda_{2} = \frac{p-2+\zeta -\theta -\yj}{\dja } \,,
    \end{equation}
     and they backreact on metric and $\varphi$ at quadratic order as
\begin{equation} \label{final2}
        \sim r^{\beta_{2}}\,,  \qquad \mathrm{where} \quad  \beta_{2} := \dja \lambda_{2} + 2\yj \,.
\end{equation}

After considering all conditions \eqref{con111}-\eqref{con7} with the conditions for \eqref{final1} and \eqref{final2}, we find that the parameter space which is shown in Table \ref{tab:5}.
\begin{table}
\centering
\begin{tabular}{|lll|}
\hline
 &  $\qquad \qquad  \qquad \qquad \quad \yj < 0$  &   \\
\hline
$\quad \  z=\yj $ &  $ \theta >p-1  $ &$ \  \zeta > 2-p+\theta-\yj, \quad \frac{\lambda_{2}}{\lambda_{1}}<\yj$ \\
\hline
 &  $\qquad  \qquad  \qquad \quad 1 < \yj $ &  \\ \hline
$\quad \ z = \yj \ \ $ & $  \theta<\dfrac{p-1}{2}\left(\yj -\sqrt{(\yj-2)^{2}+\frac{4(\yj-1) }{p-1}}\right)$ & $ \   \zeta < 2-p+\theta-\yj, \quad \frac{\lambda_{2}}{\lambda_{1}}>\yj $ \\ \hline
\end{tabular}
\caption{Parameter range: Class II-i, $V_0 >0$} \label{tab:5}
\end{table}

\subsubsection{Class II-ii: irrelevant current}

We assume that the classical solutions are written as
\begin{equation} \label{class1sol}
\begin{split}
& \dd s^2=r^{\frac{2\theta}{p-1}}\left[-\frac{\dd t^2}{r^{2z}}+\frac{L_r^{2}\dd r^2
}{r^2}+\frac{\sum\limits_{i=1}^{p-2} \dd x_{i}^2}{r^2}+\frac{\dd y^2}{r^{2\yj}}\right]\,,   \\
& {\varphi}={\dja}\log r \  \,,  \quad  A=0 \,,  \quad \chi_{i}=0 \,, \quad \chi_{p-1} = k_2 y\,.
  \end{split}
\end{equation}
By the equations of motion, the `exponent' solution-parameters ($z, \yj, \theta$) may be expressed in terms of action-parameters ($\delta,  \lambda_1 $) as
    \begin{equation}
            z = 1  \,, \qquad
            \theta = \frac{(1-p) \left(\lambda _2+(p-1)\delta\right)\delta}{2+\lambda _2^2-(p-1)\delta ^2} , \qquad \yj = \frac{\lambda _2 \left(\lambda _2+ (p-1)\delta \right)}{2+\lambda _2^2 - (p-1)\delta ^2} \,.
    \end{equation}
Note that $\lambda_1$ and $\gamma$ does not contribute to the IR solution because $k_1=a_0=0$. 
The first equation can be understood by \eqref{jjj11} where we set $k_1=0$.
The `coefficient' parameters are solved as
    \begin{align}
        &  \dja = -\frac{2\yj}{\lambda_{2}} \,,  \\
        & L_r^{2} =  \frac{(p-1-\theta) (p-1-\theta +\xi)}{V_0} \,, \\
        &k_{2}^{2} = \frac{2 V_0(\yj -1 ) }{\theta -(p-1)}    \,. 
    \end{align}
It can be understood by \eqref{jjj11} by plugging $z=1$.

The action-parameters ($\delta, \lambda_2 $) may be written in terms of solution-parameters ($\theta, \yj$):

\begin{equation} \label{actionmetric1}
\begin{split}
 \delta=\frac{2\theta}{(p-1)\dja}\,,  \quad \lambda_{2} = -\frac{2\yj}{\dja}\,, 
  \end{split}
\end{equation}
where 
\begin{equation} \label{kappa1}
\begin{split}
 \dja^{2}&=  2 \theta  \left(\frac{\theta }{p-1}-1\right)-2 (\yj -1) \yj \,,
 \end{split}
\end{equation}

Note that all formulas so far are independent of $\lambda_1$ and $\gamma$ because one axion ($k_1 x_i$) and the current are irrelevant. By turning on the subleading gauge field mode
\begin{equation}
A_{t}(r) = \, a_0 r^{ \zeta-1}  \,,
\end{equation}
we find
\begin{equation} \label{zeta14}
        \zeta = \frac{2+\gamma  \dja +3 \theta -\yj +p (p-3-\gamma  \dja -\theta +\yj )}{p-1} \,,
\end{equation}
where $\zeta$ is a function of a free action parameter $\gamma$.
This gauge field mode backreact on metric and $\varphi$ at quadratic order as
\begin{equation} \label{final11}
        \sim r^{\beta_{1}}\,,  \qquad \mathrm{where} \quad  \beta_{1} := p-2+\zeta -\theta +\yj \,.
\end{equation}
If we turn on the subleading axion mode ($k_1 x_i$) we find
    \begin{equation}
        \lambda_{1} =\frac{p-4+\zeta -\theta +\yj }{\dja } \,,
    \end{equation}
     and they backreact on metric and $\varphi$ at quadratic order as
\begin{equation} \label{final22}
        \sim r^{\beta_{2}}\,,  \qquad \mathrm{where} \quad  \beta_{2} := \dja \lambda_{1} + 2 \,.
\end{equation}

After considering all conditions \eqref{con111}-\eqref{con7} with the conditions for \eqref{final11} and \eqref{final22}, we find that the parameter space which is shown in Table \ref{tab:6}.
\begin{table}
\centering
\begin{tabular}{|lll|}
\hline
 &  $\qquad \qquad  \qquad \qquad \quad \yj \leq 0$  &   \\
\hline
$\quad \  z= 1  $& $\theta <\yj (p-1) $& $\zeta < 2-p+\theta-\yj, \quad \frac{\lambda_{1}}{\lambda_{2}}<\yj^{-1} $\\
\hline
 & \ {$ \qquad \qquad \qquad \ \   0<\yj<1$}  & \\ \hline
$\quad \  z=1$ & $\theta<\dfrac{p-1}{2}\left(1 -\sqrt{1+\frac{4(\yj-1) \yj}{p-1}}\right) $ & $ \zeta < 2-p+\theta-\yj, \quad \frac{\lambda_{1}}{\lambda_{2}}>\yj^{-1} $  \\ \hline
\end{tabular}
\caption{Parameter range: Class II-ii, $V_0 >0$} \label{tab:6}
\end{table}

\section{Thermal diffusion and butterfly velocity}  \label{sec:4}

In this section we consider diffusion in the anisotropic system.
Diffusion in strongly correlated systems is a very interesting subject because of its proposed relation to the chaos properties such as the Lyapunov time ($\tau_L$) and the butterfly velocity ($v_B$), which are introduced in \eqref{chaos111}.
At finite density, charge and energy diffusion are coupled and two diffusion constants $D_\pm$ describing the coupled diffusion of charge and energy can be obtained by the generalized Einstein relation~\cite{Hartnoll:2014lpa}\footnote{The conductivities may be diagonalized as in \cite{Davison:2015bea}.}.
\begin{eqnarray}
D_+ D_- & = & \frac{\sigma}{\chi} \frac{\kappa}{c_\rho}   \,,  \label{DD3} \\
D_+ + D_- & = & \frac{\sigma}{\chi} + \frac{\kappa}{c_\rho}  \label{DD4}
+ \frac{T(\tilde{\zeta} \sigma - \chi \alpha)^2}{c_\rho \chi^2 \sigma} \,,
\end{eqnarray}
where  $\sigma, \alpha$, and $\kappa$ are the electric, thermoelectric and thermal conductivity respectively.
$\chi$ is the compressibility, $c_\rho$ is the specific heat at fixed charge density and $\tilde{\zeta}$ is the thermoelectric susceptibility.

If the charge density is zero, since $\alpha=\tilde{\zeta}=0$, the `mixing term' (the third term in \eqref{DD4}) vanishes. In this case  $D_\pm$ are decoupled and $D_+$ and $D_-$ can be identified with the charge diffusivity ($D_C$) and the thermal diffusivity ($D_T$) respectively. The mixing term is also negligible in the incoherent regime where  momentum relaxations is strong ($k_i/\mu \gg 1,  k_i /T \gg 1$). Furthermore, It has been shown that in the low temperature limit of the scaling geometry studied in section \ref{sec:2}, the mixing term is negligible.  In this section, we consider this low temperature limit and focus on the anisotropic thermal diffusivities defined by
\begin{equation}
D_{T,x} := \frac{\kappa_{xx}}{c_\rho} \,,\qquad D_{T,y} := \frac{\kappa_{yy}}{c_\rho} \,,
\end{equation}
and a specific combinations:
\begin{equation}
\mathcal{E}_{x} := \frac{D_{T,x}}{v_{B,x}^2 \tau_L} \,,  \qquad \mathcal{E}_{y} := \frac{D_{T,y}}{v_{B,y}^2 \tau_L} \,,
\end{equation}
where $v_{B,i}$ is the butterfly velocity in the $i$ direction.

The essential idea  of the following analysis was already described in \cite{Blake:2017qgd} and we  closely follow the steps therein. Our goal here is to extend the results in \cite{Blake:2017qgd} in three aspects.  i) to understand $\mathcal{E}_x $ and $\mathcal{E}_y $  in terms of scaling exponents $z$ and  $\yj$ and ii) to identify the allowed range of $\mathcal{E}_x $ and $\mathcal{E}_y $ and see if there is any universal lower or upper bound.  iii) to extend the formalism to the case where $g_{tt} \ne g_{rr}^{-1}$.
To achieve our goals, the analysis in section \ref{sec:2} are necessary.

For simplicity let us consider $p=3$, in which case the action \eqref{EMDA} becomes
\begin{equation}
	S = \int \dd^{4}x \sqrt{-g}\left[R - \frac{1}{2} (\partial \varphi)^{2} + V(\varphi) - \frac{1}{4} Z(\varphi)F^{2} - \frac{1}{2}W_{1}(\varphi)(\partial \chi_{1})^{2}  - \frac{1}{2}W_{2}(\varphi)(\partial \chi_{2})^{2} \right].
\end{equation}
We consider a general metric solution of the form
	\begin{align} \label{gmet1}
	\dd s^2&=-D(r)\dd t^2+B(r)\dd r^2+C_{1}(r)\dd x^{2}+C_{2}(r)\dd y^{2}  \,,
	\end{align}
where we allow $B \ne D$ and slightly generalize the formulas in \cite{Blake:2017qgd} to the case $B \ne D^{-1}$.

From the metric \eqref{gmet1} the temperature and  entropy density read
	\begin{equation} \label{base1}
	T = \frac{1}{4\pi } \left. \frac{|D'|}{\sqrt{D B}}\right|_{r_{h}} \,, \qquad
	s = \left. 4\pi \sqrt{C_{1} C_{2}}\right|_{r_{h}} \,.
	\end{equation}
The electric conductivity ($\sigma_{xx}$), thermoelectric conductivity ($\alpha_{xx}$), and thermal conductivity   ($\bar{\kappa}_{xx}$) have been obtained in terms of horizon data in \cite{Donos:2014cya}.  In our convention they read\footnote{These DC formulas have been confirmed by computing the optical conductivities and taking the zero frequency limit~\cite{Kim:2014bza,Kim:2015sma,Kim:2015wba}.  See also  \cite{Blake:2013bqa, Blake:2013owa} which were the first papers developing the techniques to calculate the electric conductivity in terms of the black hole horizon data in massive gravity.}. 
%
\begin{equation} \label{conducallx}
\sigma_{xx} = \, \left.\left(\frac{\rho^2}{k_{1}^{2} W_{1} \sqrt{C_{2} C_{1}}} + Z  \sqrt{\frac{C_2}{C_{1}}}\right)\right|_{r_{h}} \,,  \quad
\alpha_{xx} = \left.\frac{4 \pi \rho}{k_{1}^{2}W_{1}}\right|_{r_{h}} \,, \quad
\bar{\kappa}_{xx} = \left.\frac{4\pi s T}{k_{1}^{2}W_{1}}\right|_{r_{h}} \,,
\end{equation}
where $\rho$ is the charge density
\begin{equation} \label{base2}
	\rho = \sqrt{\frac{C_{1} C_{2}}{D B}}Z A_t' \,,
\end{equation}
which is easily seen in \eqref{max123}. 
The thermal conductivity with an open circuit condition ($\kappa_{xx}$) is
	\begin{equation} \label{openkappa1}
	\begin{gathered}
	\kappa_{xx} = \bar{\kappa}_{xx} - \frac{T \alpha_{xx}^{2}}{\sigma_{xx}} = \left.\frac{4 \pi s T Z(\varphi(r))C_{2}(r)}{\rho^{2} + k_{1}^{2}W_{1}(\varphi(r))Z(\varphi(r))C_{2}(r)}\right|_{r_{h}}. \\
	\end{gathered}
	\end{equation}

All conductivities in \eqref{conducallx} and \eqref{openkappa1} depend on the metric, the forms of couplings and the profiles of the matter fields. However, the key observation made in \cite{Blake:2017qgd} is that the thermal conductivity with an open ciruit condition ($\kappa_{xx}$) is a function only of the metric. It can be seen from the Einstein equations.
By eliminating the second term including $V$ in \eqref{eom11} and \eqref{eom13} we have
	\begin{equation}
	\begin{split}
	  \frac{B' C_{1}'}{B^{2}} +
\frac{C_{1}'^{2}}{B C_{1}} - \frac{C_{1}' C_{2}'}{B C_{2}} - \frac{C_{1} B' D'}{B^{2} D}
	+ \frac{C_{1} C_{2} B D}{C_{2}' D'} - \frac{C_{1} D'}{B D^{2}} - \frac{2 C_{1}''}{B} + \frac{2 C_{1} D''}{B D}  \\ = 2 k_{1}^{2}W_{1} + \frac{2 Z C_{1} A_t'^{2}}{B D} \,.
	\end{split}
	\end{equation}
Note that the right hand side is a combination of the stress energy tensor in the Einstein equations and it is the combination that appears in the denominator of \eqref{openkappa1} after substituting $\rho$ with \eqref{base2}.
Thus, $\kappa_{xx}$ can be expressed only in terms of metric:
	\begin{equation} \label{kappax1}
	\kappa_{xx} = \left. \frac{32\pi^{2}T(C_{1}^{-1}C_{2})^{\frac{3}{2}} D'}{[(C_{1}^{-1}C_{2})(B D)^{-1} D'^{2}]'}\right|_{r_{h}} \,.
	\end{equation}
Note that the dependence of $\kappa_{xx}$ on the matter fields and the couplings are still implicitly encoded in the metric but there is no explicit `handle of matter' to control $\kappa_{xx}$. This suggests that there may be some universal feature.
For the conductivities in the $y$-direction,  we only need to replace the subscripts $x \rightarrow y$ and $1 \leftrightarrow 2$  from \eqref{conducallx} and \eqref{kappax1}.  For exmaple, from \eqref{kappax1}
	\begin{equation} \label{kappay1}
	\kappa_{yy} = \left. \frac{32\pi^{2}T(C_{2}^{-1}C_{1})^{\frac{3}{2}} D'}{[(C_{2}^{-1}C_{1})(B D)^{-1} D'^{2}]'}\right|_{r_{h}}.
	\end{equation}
The formula \eqref{kappax1} and \eqref{kappay1} are reduced to the ones in \cite{Blake:2017qgd} if $B = D^{-1}$.

The specific heat can be computed from the entropy density in \eqref{base1} as
\begin{equation} \label{spehe1}
 c_\rho = T \left(\frac{\partial s}{\partial T}\right)_{\rho} = \left. \frac{2\pi T(C_{1}C_{2})'}{\sqrt{C_{1}C_{2}}}\right|_{r_{h}}\frac{\partial r_{h}}{\partial T}\,,
\end{equation}
where $r_h$ is a function of $T$ obtained by the first equation of \eqref{base1}. In principle, $\frac{\partial r_h}{\partial T}$ also can be written in terms of the derivatives of the metric but it is not so illuminating.

Finally, the butterfly velocities can be computed holographically by considering a shock wave geometry and they are written in terms of the metric data at horizon~\cite{Ling:2016ibq,Blake:2017qgd}. For anisotropic case  
%
%
%

	\begin{equation} \label{butter11}
		v_{B,x} = \left. \frac{2\pi T}{\sqrt{C_{1}}m} \right|_{r_{h}}  \,, \qquad
		v_{B,y} = \left. \frac{2\pi T}{\sqrt{C_{2}}m} \right|_{r_{h}} 	\,,
		\end{equation}
where
	\begin{equation}
		m = \sqrt{\pi T \left. \left(\frac{(C_{1}C_{2})'}{C_{1}C_{2}{\sqrt{B D}}}\right) \right|_{r_{h}}}\,,
	\end{equation}
which slightly generalize the formula of \cite{Ling:2016ibq, Blake:2017qgd} to the case $B \ne D^{-1}$.

Having the general formulas for the thermal conductivity ($\kappa_{xx}, \kappa_{yy}$), specific heat ($c_\rho$), and the butterfly velocity ($v_{B,x}, v_{B,y}$) for a metric of the form \eqref{gmet1}, we turn to our anisotropic model in section \ref{sec:2}.
For all classes considered in there the metric is of the form
	\begin{equation} \label{ourmet1}
	\dd s^{2} = r^{\theta}\left( -f(r) \frac{\dd t^{2}}{r^{2z}} + \frac{L_{r}^{2} \dd r^{2}}{f(r) r^{2}} + \frac{L_{1}^{2} \dd x^{2}}{r^{2}} + \frac{L_{2}^{2}\dd y^{2}}{r^{2\yj}}\right) \,,
	\end{equation}
with the emblackening factor
\begin{equation} \label{}
f(r) =  1- \left(\frac{r}{r_h}\right)^{z+1-\theta+ \yj} \,.
\end{equation}	

The temperature \eqref{base1} is related to the horizon position $r_h$ as follows.
\begin{equation} \label{tt11}
T = \frac{r_{h}^{-z}|1+z-\theta+\yj|}{4 \pi L_{r}} \,,  \qquad r_{h} = \left(\dfrac{|1+z-\theta+\yj|}{4\pi L_{r} T}\right)^{\frac{1}{z}}\,.
\end{equation}
Thermal conductivities \eqref{kappax1} and \eqref{kappay1} are
	\begin{equation} \label{kxxyy}
	\kappa_{xx} = \frac{4 \pi }{2(1 - z)}L_{r} L_{1}^{-1}L_{2} r_{h}^{\theta - z - \yj+1} \,, \qquad
	\kappa_{yy} = \frac{4 \pi }{2(\yj - z)} L_{r} L_{1}L_{2}^{-1} r_{h}^{\theta - z + \yj-1} \,,
\end{equation}
%
	%
and the specific heat \eqref{spehe1} is computed as
	\begin{equation} \label{crho11}
	c_{\rho} = T \left(\frac{\partial s}{\partial T}\right)_{\rho}  = 4\pi  \frac{1 + \yj - \theta}{z}L_{1}L_{2} r_{h}^{\theta - 1-\yj} \,,
	\end{equation}
%
	%
from the entropy density  \eqref{base1}
	\begin{equation}
	s = 4 \pi L_{1}L_{2} r_{h}^{\theta - (1 + \yj)} \sim T^{\frac{1+\yj - \theta}{z}} \,.
	\end{equation}
In \eqref{kxxyy} and \eqref{crho11} we replaced $T$ with $r_h$ by using \eqref{tt11} to simplify the expression.

Thus, the diffusivities are	
	\begin{equation}
	\begin{split}
	D_{T,x} & = \frac{\kappa_{xx}}{c_{\rho}} = \frac{z}{(2z-2)(\theta-1-\yj )}  L_{r} L_{1}^{-2} r_{h}^{2 - z} \,, \\
	D_{T,y }& = \frac{\kappa_{yy}}{c_{\rho}} = \frac{z }{\left(2z-2(2-\yj)\right)(\theta-1-\yj)} L_{r} L_{2}^{-2} r_{h}^{2\yj - z}  \,,
	\end{split}
	\end{equation}
	%
and the butterfly velocities \eqref{butter11} are

	%
%
\begin{equation}
		v_{B,x}^2= \frac{2\pi T  }{\theta-1-\yj} L_{r} L_{1}^{-2}r_{h}^{2-z} \,, \qquad
		v_{B,y}^2  = \frac{2\pi T }{\theta-1-\yj}  L_{r} L_{2}^{-2}r_{h}^{2\yj-z} \,.
\end{equation}
%
%

Finally, by noticing that $\tau_L = (2\pi T)^{-1}  $ we have
\begin{align} \label{ab12}
& \mathcal{E}_{x} = \frac{D_{T,x}}{v_{B,x}^2 \tau_L}  =  \frac{1}{2}\frac{z_x}{z_x-1}  = \frac{1}{2}\frac{z}{z-1}  \,,  \\
& \mathcal{E}_{y} = \frac{D_{T,y}}{v_{B,y}^2 \tau_L}  = \frac{1}{2}\frac{z_y}{z_y-1}  = \frac{1}{2}\frac{z}{z-\yj} \,.
\end{align}
Notice that the $\mathcal{E}_{x}$ and $\mathcal{E}_{y}$ depend only on $z$ and $\yj$ irrespective of $\theta$ and $\zeta$. They are also independent of charge density $\rho$ and momentum relaxations $k_1$ and $k_2$. This universality is nontrivial because the thermal conductivities, specific heat and butterfly velocity, all of them depend on $(\theta, \zeta, \rho, k_1, k_2)$ through $(L_r, L_1, L_2, r_h)$.  When it comes to the combinations $\mathcal{E}_{x}$ and $\mathcal{E}_{y}$, all $L_r, L_1$, $L_2$ and $r_h$ are canceled out.

To investigate if there is any lower or upper bound of $\mathcal{E}_{x}$ and $\mathcal{E}_{y}$, we need to understand the parameter region of $z$ and $\yj$.  We will restrict ourselves to positive $z_i$.  Based on the allowed parameter region obtained in section \ref{sec:2} we find
\begin{itemize}
\item Class I and II 
\begin{align}
\frac{\lambda_{2}}{\lambda_{1}} \ge 1 \quad  \Rightarrow &  \quad \frac{1}{2} \le \mathcal{E}_x <  \frac{1}{2}\left( \frac{1}{1- {\yj^{-1}} }\right) \,, \qquad  \frac{1}{2} \le \mathcal{E}_y  \,, \label{uytr}\\
\frac{\lambda_{1}}{\lambda_{2}}  \ge 1  \quad  \Rightarrow  & \quad  \frac{1}{2} \le \mathcal{E}_x  \,, \qquad 
\frac{1}{2} \le \mathcal{E}_y <  \frac{1}{2}\left( \frac{1}{1- {\yj} }\right) \,,
\end{align}
where $\yj = \frac{\lambda_2}{\lambda_1}$.
\item Class I-i 
\begin{align}
\frac{\lambda_{2}}{\lambda_{1}} > 1 \quad  \Rightarrow &  \quad \frac{1}{2} \le \mathcal{E}_x <  \frac{1}{2}\left( \frac{1}{1- {\yj^{-1}} }\right) \,, \qquad  \frac{1}{2} \le \mathcal{E}_y  \,,
\end{align}
where $\yj = -\frac{\gamma +\delta+\lambda_{1}}{\lambda_{1}}$. Here $k_2=0$ and  $\frac{\lambda_1}{\lambda_2} > 1$ is not allowed.
\item Class I-ii 
\begin{align}
\frac{\lambda_{1}}{\lambda_{2}}  > 1  \quad  \Rightarrow  & \quad  \frac{1}{2} \le \mathcal{E}_x  \,, \qquad 
\frac{1}{2} \le \mathcal{E}_y <  \frac{1}{2}\left( \frac{1}{1- {\yj} }\right) \,,
\end{align}
where $\yj = -\frac{\lambda_{2}}{\gamma+\delta+\lambda_{2}}$. Here $k_1=0$ and  $\frac{\lambda_2}{\lambda_1} > 1$ is not allowed.
\item Class II-i 
\begin{align}
\frac{\lambda_{2}}{\lambda_{1}} > 1 \quad  \Rightarrow &  \quad \frac{1}{2} \le \mathcal{E}_x =  \frac{1}{2}\left( \frac{1}{1- {\yj^{-1}} }\right)  \,,
\end{align}
where  $\yj  = \frac{2 - 2 \delta^{2} + \lambda_{1}^{2}}{\lambda_{1}(2 \delta + \lambda_{1})}$. Here $k_2=0$ and $\frac{\lambda_1}{\lambda_2} > 1$ is not allowed. $z=\yj$ so $\mathcal{E}_y$ is not computed in our method.
\item Class II-ii 
\begin{align}
\frac{\lambda_{1}}{\lambda_{2}}  > 1  \quad  \Rightarrow  & \quad  
\frac{1}{2} \le \mathcal{E}_y =  \frac{1}{2}\left( \frac{1}{1- {\yj} }\right)  \,,
\end{align}
where $\yj = \frac{\lambda_{2}(2 \delta + \lambda_{2})}{2 - 2 \delta^{2} + \lambda_{2}^{2}}$ with $z = 1$. Here $k_1=0$ and  $\lambda_2/\lambda_1 > 1$ is not allowed.  $z=1$ so $\mathcal{E}_x$ is not computed in our method.
\item Class III
\begin{align}
\frac{\lambda_{1}}{\lambda_{2}}  = 1  \quad  \Rightarrow  & \quad  
\frac{1}{2} \le \mathcal{E}_x  = \mathcal{E}_y = \frac{1}{2}\left(\frac{z}{z-1}\right)\,. \label{thth}
\end{align}
\item Class IV: $z=1$ so $\mathcal{E}_i$ cannot be computed in our method. 
\end{itemize}

We find that the lower bound of $\mathcal{E}_i$ is always $1/2$. However, contrary to the isotropic case, there may be an upper bound for class I, II, I-i, and I-ii. All of these have at least one marginally relevant axion. This upper bound can be understood from the fact  that for $0 < \yj \le 1$, $z > 1$ and for $\yj \ge 1$, $z > \yj$.  For example, in \eqref{uytr}, even though $\mathcal{E}_{x}$ does not depend on $\yj$ explicitly, its range depends on $\yj$ because the available parameter range of $z$ depends on $\yj$. In this case it is $z > \yj$, which gives an upper bound.

%

%
	%
%
%

\section{Conclusion}  \label{sec:5}

In this paper, we have studied the holographic systems so called `Q-lattice' or Einstein-Maxwell-Dilaton theory coupled to `Axion' fields (EMDA).  The dilaton is introduced to support the scaling IR geometry and  the axion fields are included to break translational symmetry. Our main focus is to study the effect of spatial anisotropy which is introduced  in two ways: i) by making the different dilaton couplings  to axion fields and ii) by considering the different momentum relaxation parameters for spatial directions. The former is characterized by  $\lambda_1$ and $\lambda_2$ in \eqref{IRcoupling} and the latter is done by  $k_1$ and  $k_2$ in \eqref{backmaster}.

First, we have extended four classes of the isotropic IR geometry \cite{Gouteraux:2014hca} to the anisotropic case, which yields eight classes.
For marginally relevant axion, where the momentum relaxation parameters ($k_1, k_2$) appear explicitly in the leading IR solution, the anisotropy of $k_i$ and $\lambda_i$ are related ($i=1,2$). However, for irrelevant axion, where the momentum relaxation parameters ($k_1, k_2$) do not appear in the leading IR solution, the sub-leading order mode analysis imposes the conditions $\lambda_1=\lambda_2$, and $k_i$ and  $\lambda_i$ are not related. It is also possible that only one of $k_i$ is zero. Therefore, in total there are four classes in terms of `relevance' of axions: in the leading order solution, both $k_i$ are nonzero, both $k_i$ are zero, only one $k_i$ is zero (in 2+1 field theory dimension, $k_1=0$ and $k_2=0$ are equivalent.)
For every classes the current may be marginally relevant or irrelevant. i.e. the temporal gauge filed $A_t$ may be non-zero or zero in the leading order solution. Therefore, we have eight classes in total. 

The solutions have many parameters so called `solution-parameters', which include two critical exponents: $z_i$ for a dynamical exponent along $i$-direction, $\theta$ for a hyperscaling violating exponent, and $\zeta$ for the anomalous dimension of the field theory charge density. In our representation $z_x = z$ and $z_y = z/\yj$ where  $\yj$ characterizes the anisotropy between $x$-direction and $y$-direction. These solution-parameters should be restricted by some physical conditions such as reality, positive specific heat, and null energy conditions. We have identified those conditions in Tables \ref{tab:1}-\ref{tab:6} and Fig. \ref{fig:1}.

Next, we have considered thermal diffusion in anisotropic cases.
For the holographic systems with the metric
	\begin{align}
	\dd s^2&=-D(r)\dd t^2+B(r)\dd r^2+C_{1}(r)\dd x^{2}+C_{2}(r)\dd y^{2}  \,,
	\end{align}
the thermal conductivity, specific heat, butterfly velocity in $x$-direction can be computed in terms of the horizon data as
	\begin{align} \label{ }
	& \kappa_{xx} = \left. \frac{32\pi^{2}T(C_{1}^{-1}C_{2})^{\frac{3}{2}} |D'|}{[(C_{1}^{-1}C_{2})(B D)^{-1} D'^{2}]'}\right|_{r_{h}} \,, \\
	& c_\rho = \left. \frac{2\pi T(C_{1}C_{2})'}{\sqrt{C_{1}C_{2}}}\right|_{r_{h}}\frac{\partial r_{h}}{\partial T}\,,  \qquad v_{B,x}^2 = \left.   \frac{4\pi T(C_{2}\sqrt{B D})}{(C_{1}C_{2})'} \right|_{r_{h}} \,,
	\end{align}
which give
\begin{equation} \label{hhh1}
\mathcal{E}_x=\frac{D_{T,x}}{v_{B,x}^{2} \tau_L}  = \left. \frac{8 \pi C_{1}^{-1}C_{2}|D'|}{\sqrt{BD}[C_{1}^{-1}C_{2}(BD)^{-1}D'^{2}]'}\right|_{r_{h}}\frac{\partial T}{\partial r_{h}} \,,
\end{equation}
where $\tau_L = (2\pi T)^{-1}$ is the Lyapunov time. We may obtain the quantities in  $y$-direction by switching the subscript $1 \leftrightarrow 2$.
Notice that $\kappa_{xx}$ is originally a function of the couplings and the profiles of the matter fields. All these explicit matter dependences were replaced by the metric thanks to the Einstein equations, which suggests  that there may be some universal feature. However, \eqref{hhh1} is still a complicated function of the metric components so can not guarantee a universality by itself. For example, in general, it can be a function of $r_h$, which is a function of temperature, charge density, and momentum relaxations etc.

For the IR scaling geometry we have studied in section \ref{sec:2} the combination \eqref{hhh1} is reduced to a simple universal form:
\begin{equation} \label{nnmmb}
\mathcal{E}_{x}  = \frac{1}{2} \left(\frac{z_x}{z_x-1}\right) = \frac{1}{2} \left(\frac{z}{z-1} \right) \,, \qquad \mathcal{E}_{y}   =  \frac{1}{2} \left( \frac{z_y}{z_y-1} \right)=  \frac{1}{2} \left( \frac{z}{z-\yj}  \right)\,.
\end{equation}
Notice that in this geometry the parameters $\mathcal{E}_{x}$ and $\mathcal{E}_{y}$ depend only on $z$ and $\yj$ irrespective of $\theta$, $\zeta$, charge density $\rho$ and momentum relaxations ($k_1$ and $k_2$). This universality is due to cancellations between
three quantities $\{\kappa_{xx} (\kappa_{yy}), c_\rho, v_{B,x} (v_{B,y})\}$, all of which depend on $\{\theta, \zeta, \rho, k_1, k_2\}$ as well as $\{z, \yj\}$.  

We also studied the possible range of $\mathcal{E}_{x}$ and $\mathcal{E}_{y}$ to see if there is any universal lower or upper bound. Based on the parameter range analyzed in section \ref{sec:2} we find that the lower bound of $\mathcal{E}_i$ is always $1/2$. However, there may be an upper bound due to anisotropy, which was summarized in \eqref{uytr}-\eqref{thth}. It would be interesting to understand how much this lower and upper bound is robust in deformation of the theory, for example, with finite magnetic field~\cite{Blake:2017qgd, Amoretti:2016cad} or with higher derivative gravity in `Q-lattice' models.

In holography, it might be possible to construct theories with a less (or non) universal $\mathcal{E}_i$ or without any universal bound of  $\mathcal{E}_i$ by considering some complicated enough bulk models. In condensed matter systems, as pointed out in \cite{Blake:2017qgd}, the expression $\mathcal{E}_i$ in \eqref{nnmmb} is not expected to be universal for all systems with the same dynamical critical exponent. For example, some models with $z = 3/2$ may give $\mathcal{E} \sim 0.42$ which is different from \eqref{nnmmb}~\cite{Patel:2016aa}. 
Thus, a counter example regarding the universality in holographic models is not always bad. The important direction will be to classify the conditions for the universality and understand its origin from both gravity and condensed matter perspective, towards experimental understanding and applications.

\acknowledgments
We would like to thank Matteo Baggioli, Blaise Gouteraux, Ki-Seok Kim, Andrew Lucas, Sang-Jin Sin and Yunseok Seo for valuable discussions and correspondence.  The work of D. Ahn, Y. Ahn, H.-S. Jeong, K.-Y. Kim and C. Niu was supported by
Basic Science Research Program through the National Research Foundation of Korea(NRF) funded by the Ministry of Science, ICT $\&$ Future Planning(NRF- 2017R1A2B4004810) and GIST Research Institute(GRI) grant funded by the GIST in 2017. The work of W. J. Li was supported by the Fundamental Research Funds for the Central Universities No. DUT 16 RC(3)097 as well as NSFC Grants No. 11375026.
We also would like to thank the APCTP(Asia-Pacific Center for Theoretical Physics) focus program,``Geometry and Holography for Quantum Criticality'' in Pohang, Korea for the hospitality during our visit,
where part of this work was done.

\appendix

\section{Consistency check by coordinate transformation}

In order to compare the results in section \ref{sec:4} with \cite{Blake:2017qgd}, let us summarize the main formulas in \cite{Blake:2017qgd}.   The metric is written as
\begin{equation} \label{compa1}
		\dd s^{2} =  -U(\tilde{r})\dd \tilde{t}^{2} + \frac{\dd \tilde{r}^{2}}{U(\tilde{r})} + h_{1}(\tilde{r}) \dd \tilde{x}^{2} + h_{2}(\tilde{r}) \dd \tilde{y}^{2} \,,
\end{equation}
of which IR geometry is parameterized as
\begin{equation}
	U(\tilde{r}) = L_{t}^{-2} \tilde{r}^{u_{1}}\left(1 - \frac{\tilde{r}_{h}^{\Delta}}{\tilde{r}^{\Delta}}\right), \quad h_{1}(\tilde{r}) = L_{x}^{-2}\tilde{r}^{2v_{1}}, \quad h_{2}(\tilde{r}) = L_{y}^{-2}\tilde{r}^{2v_{2}} \,,
\end{equation}
where $\Delta = v_{1} + v_{2} + u_{1} - 1$.
The parameters $\mathcal{E}_x $ and $\mathcal{E}_y $ are obtained as
\begin{equation}
	\mathcal{E}_x =  \frac{D_{T,x}}{v_{B,x}^2 \tau_L} = \frac{u_{1} - 1}{u_{1} - 2v_{1}}, \qquad \mathcal{E}_y  = \frac{D_{T,y}}{v_{B,y}^2 \tau_L} = \frac{u_{1} - 1}{u_{1} - 2v_{2}} .
\end{equation}

As a consistency check, we may consider the coordinate transformation between \eqref{compa1}  and our metric \eqref{ourmet1}
\begin{equation}
\dd s^{2} = r^{\theta}\left( -f(r) \frac{\dd t^{2}}{r^{2z}} + \frac{L_{r}^{2} \dd r^{2}}{f(r) r^{2}} + \frac{L_{1}^{2} \dd x^{2}}{r^{2}} + \frac{L_{2}^{2}\dd y^{2}}{r^{2\yj}}\right) \,. \nonumber
\end{equation}
Two metrics are related as follows.
	\begin{equation}
		\begin{split}
		&\{\tilde{t}, \tilde{r}, \tilde{x}, \tilde{y} \} = \left\{t, \frac{L_{r}}{\theta - z} r^{\theta - z}, x, y \right\}\,,  \\
			&u_{1} = \frac{\theta - 2z}{\theta - z}, \qquad 2v_{1} = \frac{\theta - 2}{\theta - z},\qquad 2v_{2} = \frac{\theta - 2\yj}{\theta -z}\,, \\
			 &L_{t}^{2} = \left(\frac{L_{r}}{\theta - z}\right)^{\frac{2z - \theta}{z - \theta}}\,, \quad
			L_{x}^{2} = \frac{1}{L_{1}^{2}}\left(\frac{L_{r}}{\theta - z}\right)^{\frac{\theta - 2}{\theta - z}}\,, \qquad 		L_{y}^{2} = \frac{1}{L_{2}^{2}}\left(\frac{L_{r}}{\theta - z}\right)^{\frac{\theta - 2\yj}{\theta - z}} \,.
		\end{split}
	\end{equation}
We have confirmed all of our results agree to \cite{Blake:2017qgd} by using this coordinate transformation. For example,
\begin{equation}
	\mathcal{E}_x  = \frac{z}{2z-2} = \frac{u_{1} - 1}{u_{1} - 2v_{1}}, \qquad \mathcal{E}_y = \frac{z}{2z-2
\yj} = \frac{u_{1} - 1}{u_{1} - 2v_{2}} .
\end{equation}

\bibliographystyle{JHEP}

\providecommand{\href}[2]{#2}\begingroup\raggedright\endgroup

\end{document}